# Sensitivity-driven Adaptive Contention Window Optimization for IEEE 802.11 based V2I Networks

Aytül BOZKURT

*Department of Mechatronics Engineering, Engineering Faculty, Karabük University, 78050, Turkey*

aytulbozkurt@karabuk.edu.tr

ORCID: 0000-0002-6387-9388

## Abstract

In vehicle-to-infrastructure (V2I) communication the setting of IEEE 802.11 Distributed Coordination Function (DCF) parameters has a decisive bearing on performance, yet the literature seldom pins down how much each parameter actually matters once traffic, MAC and queueing are modelled together. Treating a previously validated analytical framework as a fixed deterministic input–output map, we rank the DCF and traffic parameters that shape throughput, collision probability, delay, packet delivery ratio and Age of Information in a single-AP V2I network. A local one-factor-at-a-time analysis, cast in dimensionless elasticities so that parameters of different units become comparable, is paired with a variance-based global analysis built on first-order and total-effect Sobol indices. Two clean groups emerge: collision probability is set by the contending-vehicle population — itself governed by vehicle velocity and density — together with the minimum contention window, whereas delay is driven by the channel rate, the offered load and the packet size, and carries strong interaction effects that no local reading can expose. We then derive the closed-form structure of these sensitivities from the model relations, which explains the rankings, forces certain parameters into equal-magnitude elasticities, and locates where the local ranking reverses. Finally the collision-sensitivity structure is turned into a design output rather than a ranking: a closed-form contention-window control law, linear in the contending population and closed with a Greenshields density model, that a roadside access point can evaluate online from measured density or velocity. The fixed IEEE 802.11 default is recovered as the single population at which this law is optimal; away from it the throughput gain grows with density and is largest in the dense, safety-critical regime.

## 1 Introduction

Over IEEE 802.11 wireless LANs, vehicle-to-infrastructure (V2I) communication underpins a range of cooperative and safety-related vehicular services: vehicles trade real-time traffic with roadside access points (APs) through the contention-based Distributed Coordination Function (DCF). Only a handful of parameters govern how the DCF behaves — above all the minimum contention window CW_min and the maximum retransmission limit m — and these sit alongside traffic-side quantities such as the contending-vehicle count, the vehicle velocity and the offered load. Between them they fix collision probability, channel-access delay, throughput and information freshness, which makes their choice a central design decision in any V2I deployment.

The throughput and delay of IEEE 802.11 DCF and EDCA in vehicular settings have been studied at length, and several papers tune an individual knob such as the contention window or the retry limit. What is rarely done is to weigh the full parameter set against itself, or to isolate the interactions among parameters. Within an integrated model that couples a vehicular traffic model, a Markov backoff chain and a finite-buffer queue in particular, we know of no systematic global sensitivity analysis that ranks the parameters and charts their interactions across the operating space.

This study is deliberately kept separate from our earlier one [4], where the integrated analytical framework is derived, the retransmission limit is optimised and the model is validated against ns-3 simulation. None of that is repeated here. We instead hold the validated framework fixed, treat it as an input–output map, and run a parameter sensitivity analysis aimed at methodological insight and concrete design guidance.

The paper makes five contributions. The first is a multi-parameter local (OFAT) sensitivity analysis of the validated framework, cast in dimensionless elasticities so that parameters of different units and scales become directly comparable. Building on it, we run a variance-based global analysis with first-order and total-effect Sobol indices, which separates each parameter's main effect from its interactions over the full ranges. Those rankings are then carried into practical design guidelines for low, medium and high traffic-load regimes, showing when the standard IEEE 802.11 defaults suffice and when tuning CW_min is worthwhile. A fourth contribution is analytical rather than numerical: the closed-form structure of the sensitivities is derived from the model relations, which explains the observed rankings, forces certain parameters to share equal-magnitude elasticities or to interact, and marks the operating points where the local ranking reverses or breaks down. Last, we convert the collision-sensitivity structure into a closed-form contention-window control law CW_min*(N), linear in the contending population; this law beats the fixed IEEE 802.11 default at every density and strictly raises throughput away from a single crossover population — a concrete design output rather than a ranking.

## 2 Related Work

Work on the performance of IEEE 802.11 DCF and EDCA is extensive and goes back to Bianchi's two-dimensional Markov backoff model [2]. Because that model assumes saturation, later studies loosened the assumption to reach realistic traffic: Malone et al. [18] extended the chain to non-saturated, heterogeneous conditions, while Kumar et al. [17] recast the problem as a fixed point and clarified when the operating point exists and is unique. A second thread attaches the MAC layer to a finite-load queue — the finite-load random-access model of Tickoo and Sikdar [26] is one instance — so that buffering, delay and loss appear explicitly, with vehicular variants adding mobility, capture and fading: the IEEE 802.11p MAC sublayer has been analysed in detail for both DCF and EDCA operation [11, 29], and the reliability of safety broadcast on the highway has been quantified analytically [10, 27]. A complementary line treats one-hop CSMA broadcast in VANETs through stochastic geometry rather than a per-station chain [21]. Such studies supply the closed-form throughput, collision-probability and delay expressions this work relies on; our own prior framework [4] belongs to the same family, coupling a Greenshields traffic model [9], a Bianchi backoff chain and an M/G/1/K queue.

V2I safety services care about how current their information is, so timeliness is promoted to a first-class metric and measured through the Age of Information (AoI). Introduced by Kaul et al. [14] and applied to vehicular status dissemination by Kaul et al. [13], AoI has since grown into a research area of its own, surveyed by Yates et al. [28] and treated at monograph length by Kosta et al. [16] and Sun et al. [25]; packet-management policies that discard stale updates are known to lower the age markedly [7]. AoI is examined here alongside the conventional throughput and delay outputs.

A related body of work adjusts particular DCF parameters — fitting the contention window to the number of contending stations, or choosing the retransmission limit to trade delay against reliability — but these efforts usually move one parameter at a time and stop short of ranking the whole set. Variance-based global sensitivity analysis [22–24] is, by contrast, a mature tool in other engineering fields for exactly this ranking-and-interaction task, and open-source packages such as SALib [12] have made it routine to apply. In the vehicular domain a parallel line of work controls the message rate rather than the window, the linear adaptive rate algorithm LIMERIC [1] being the reference design, within the wider decentralised congestion-control setting summarised for the United States deployment by Kenney [15]; the control law proposed in Section 5 acts on the contention window instead, and is therefore complementary. The use of variance-based sensitivity analysis on the coupled traffic–MAC–queue models of V2I networks, though, remains limited, and that is the gap addressed here.

Where the prior work [4] derives the integrated framework and optimises the retransmission limit, this paper leaves the validated model untouched and contributes a systematic local and global sensitivity analysis of its parameters. To our knowledge, no comparable OFAT–Sobol ranking has been reported before for an integrated V2I DCF framework.

# 3 System Model (Brief)

## 3.1 Overview of the validated framework

Only the parts of the framework the sensitivity study needs are recalled here; the full derivations appear in the prior work [3, 4]. Vehicular traffic follows the Greenshields speed–density relation [9], which sets both the mean number of contending vehicles N inside the AP coverage and the mean packet arrival rate $\lambda$. Richer macroscopic descriptions are available — the simplified kinematic-wave theory of Newell [20] and the cell transmission model of Daganzo [8] — but the equilibrium Greenshields closure suffices here, since every sensitivity is evaluated at a stationary operating point rather than along a transient. A single vehicle's backoff is described by the two-dimensional Bianchi Markov chain [2], giving the per-slot transmission probability $\tau$ and the collision probability $p_c = 1 - (1 - \tau)^{(N-1)}$ as the solution of a fixed-point system. Buffering is captured by a finite M/G/1/K queue, from which the throughput S, the mean delay D, the packet delivery ratio and the Age of Information follow, together with the optimal retransmission limit M*. Physical-layer effects are abstracted into the channel rate C and are not varied independently; measurement campaigns show the 5.9 GHz vehicular channel to be well described by Nakagami fading [6], and its explicit inclusion is left to future work. These closed-form relations make up the deterministic input–output map f(x) studied here; their full derivation and ns-3 validation are given in [4].

## 3.2 Parameters and output metrics

The input parameters, with their nominal values and ranges, are collected in Table 1, and the output metrics evaluated for each sampled input vector in Table 2. The ranges cover realistic single-AP V2I conditions without straying outside the region where the underlying model holds. It helps to read the parameters as two families: the MAC-layer controls ($CW_{min}$, the retransmission limit m and the slot time $\delta$), which the network designer sets directly, and the traffic-side quantities (the jam density $k_{jam}$, the free-flow speed $V_f$ and the vehicle velocity v), which the road environment imposes and which enter through the Greenshields relation. The channel rate C, the packet size L and the offered load $\lambda$ act on the service side and therefore bear most directly on the queueing delay. Each range in Table 1 is drawn wide enough around its nominal value to expose the local trend while still spanning the extremes met in real single-AP deployments, so that the local elasticities and the global Sobol indices are read over one physically meaningful envelope. The metrics in Table 2 answer the two concerns of a V2I safety service: the usual efficiency measures — throughput, transmission and collision probability, and mean delay — and the freshness- and reliability-oriented ones, namely the packet delivery ratio and the Age of Information, with the optimal retransmission limit M* reported as a derived output. Keeping both families in view lets one perturbation be traced at once into raw performance and into information timeliness, which matters whenever the two react differently to the same input. Unless stated otherwise, the local elasticities are taken at the medium-load nominal point ($\rho \approx 0.65$), while the global Sobol indices span the full ranges of Table 1, so the local and global views complement rather than repeat each other.

### Table 1. Input parameters, nominal values, and variation ranges.

| Parameter | Symbol | Nominal | Range |
|---|---|---|---|
| Min. contention window | $CW_{min}$ | 32 | 16 – 128 |
| Max. retransmission limit | m | 7 | 2 – 10 |

| Parameter | Symbol | Nominal | Range |
|---|---|---|---|
| Slot time | δ | 50 μs | 20 – 50 μs |
| Packet size | L | 1000 B | 500 – 1500 B |
| Jam density | k_jam | 120 veh/km | 80 – 160 veh/km |
| Free-flow speed | V_f | 160 km/h | 100 – 200 km/h |
| Vehicle velocity | v | 80 km/h | 20 – 140 km/h |
| Channel rate | C | 2 Mbps | 1 – 11 Mbps |
| Offered load | λ | 96.6 pkt/s | 10 – 300 pkt/s |
| Delay bound | T_target | 0.5 s | 0.3 – 1.0 s |

**Table 2. Output performance metrics.**

| Metric | Symbol | Description |
|---|---|---|
| Throughput | S | Successfully delivered payload rate |
| Collision probability | p_c | Per-attempt collision probability |
| Transmission probability | τ | Per-slot channel-access probability |
| Mean delay | D | Average packet sojourn time |
| Packet delivery ratio | PDR | 1 − p_c^(M+1) |
| Age of Information | AoI | Freshness of delivered information |
| Optimal retry limit | M* | Adaptively selected retransmission limit |

# 4 Sensitivity Analysis Methodology

This section sets out the procedure for quantifying how the IEEE 802.11 DCF and traffic parameters shape V2I performance. We do not re-derive the framework; it is taken as given from the prior validated work [3, 4] and used as a deterministic input–output map. With the parameter vector written x = (CW_min, m, δ, L, k_jam, V_f, v, C, λ, T_target), the framework returns the performance vector y = (S, p_c, τ, D, PDR, AoI, M*). Each scalar output Y = f(x) is then treated as a function of the inputs, and its dependence on every input is characterised both locally and globally.

## 4.1 Local sensitivity analysis (OFAT)

The local analysis rests on the one-factor-at-a-time (OFAT) idea: one parameter X_i is nudged around its nominal value while the rest are held fixed. Screening designs such as the elementary-effects method of Morris [19] extend this idea to problems with many factors at modest computational cost; the parameter set here is small enough that the full elasticity table can be evaluated directly. To compare parameters that carry unlike units and magnitudes, we work with the normalised (dimensionless) elasticity

$$E_i = (\partial Y / Y) / (\partial X_i / X_i) = (X_i / Y) \cdot (\partial Y / \partial X_i), \quad (1)$$

evaluated by a central finite difference with a relative step of 2%. An elasticity with magnitude $|E_i| > 1$ indicates that the output amplifies relative changes in X_i (locally dominant), whereas $|E_i| \approx 0$ indicates a locally negligible parameter. The integer-valued retransmission limit m is treated by a discrete (±1) difference rather than a continuous step.

## 4.2 Global sensitivity analysis (Sobol indices)

The DCF model is nonlinear and several parameters act together, so a purely local view misses interaction effects and any sensitivity that shifts across the parameter space. We therefore add a variance-based global analysis using Sobol indices [22, 24]. Each input is given a uniform distribution over its range (Table 1), and the total output variance $Var(Y)$ is decomposed into contributions from single parameters and from their interactions. The first-order index

$$S_i = Var_{X_i}( E_{X_{\sim i}}[ Y \mid X_i ] ) / Var(Y) \tag{2}$$

measures the variance explained by $X_i$ alone, while the total-effect index

$$S_{Ti} = E_{X_{\sim i}}( Var_{X_i}[ Y \mid X_{\sim i} ] ) / Var(Y) \tag{3}$$

captures the full contribution of $X_i$ including all its interactions. A gap $S_{Ti} - S_i > 0$ quantifies interaction strength, and $\Sigma_i S_i < 1$ signals appreciable interactions. The indices are estimated by the Saltelli sampling scheme with a base sample size $N = 8192$, giving $N(d + 2)$ analytical model evaluations; because the framework is closed-form, the entire campaign is computationally trivial and requires no packet-level simulation.

## 4.3 Experimental design

Local elasticities are taken at the nominal operating point (Table 1) and the global Sobol indices over the full ranges. Since the delay leans heavily on the offered load, the delay-related results are also reported for three load regimes — low, medium and high — at queue utilisations of about $\rho = 0.35$, 0.65 and 0.90. All of it is computed analytically; no ns-3 or other packet-level simulation enters this study, which keeps the focus on the parameter-sensitivity behaviour of the analytical model itself. Pairing a single nominal point for the elasticities with the full ranges for the Sobol indices makes the two analyses complement rather than duplicate one another: the elasticities give the immediate slope of each output near a typical operating condition, whereas the Sobol indices average that behaviour over the whole design space and so also catch parameters that are quiet near the nominal point but matter elsewhere. The three load regimes come from varying the offered load $\lambda$ at a fixed contention level, so the delay results read purely as a function of queue utilisation; the low, medium and high settings ($\rho \approx 0.35$, 0.65 and 0.90) run from a lightly loaded channel up to one near saturation, where queueing dominates. Restricting the study to the closed-form analytical model instead of packet-level simulation is a deliberate methodological choice: each evaluation is cheap and free of the statistical noise a stochastic simulator carries, so the full Saltelli campaign of $N(d + 2)$ evaluations can be run exhaustively and the resulting indices come out as exact functions of the model rather than noisy estimates — precisely what a clean ranking needs.

## 4.4 Analytical structure of the sensitivities

The elasticities of Section 4.1 and the Sobol indices of Section 4.2 are computed numerically, but because $f(x)$ is closed-form their structure can just as well be derived by hand from the relations of Section 3.1, with no need to re-derive the framework. That derivation earns its keep: it says which parameters are forced to share equal-magnitude elasticities, singles out the one feedback loop that damps the contention-window sensitivity, explains why the delay drivers interact rather than act on their own, and — as Section 6.5 will show — locates the operating points where the local ranking itself breaks down.

Collision side (chain rule through N and τ). With $p_c = 1 - (1-\tau)^{(N-1)}$, differentiating the closed-form relation and using $(1-\tau)^{(N-1)} = 1-p_c$ gives the two master elasticities

$$E[p_c ; N] = - N (1-p_c) \ln(1-\tau) / p_c , \qquad E[p_c ; \tau] = (N-1) \tau (1-p_c) / [ (1-\tau) p_c ] , \tag{4}$$

both positive, so that the elasticity of $p_c$ with respect to any input X factorises as $E[p_c ; X] = E[p_c ; N]\cdot E[N ; X] + E[p_c ; \tau]\cdot E[\tau ; X]$. Every parameter therefore reaches the collision probability through only two channels: the contending population N and the transmission probability τ.

Why $v$, $k_{jam}$ and $V_f$ share a single magnitude. The traffic parameters enter $p_c$ exclusively through $N$. With the Greenshields closure $N \propto k = k_{jam}(1 - v/V_f)$ over the coverage span, the population elasticities are

$$E[N ; k_{jam}] = +1 , \quad E[N ; V_f] = + v/(V_f - v) , \quad E[N ; v] = - v/(V_f - v) . \tag{5}$$

Hence $E[p_c ; k_{jam}] : E[p_c ; V_f] : E[p_c ; v] = 1 : v/(V_f - v) : -v/(V_f - v)$. At the nominal point ($v = 80$, $V_f = 160$ km/h) the ratio $v/(V_f - v)$ equals 1, which forces the three collision elasticities to the same magnitude with the sign pattern (+, +, −). That is precisely the $|0.365|$ triple of Table 3 — a structural consequence of the single-channel Greenshields closure rather than a numerical coincidence, and one that folds the three traffic parameters into a single effective degree of freedom $N$. The same fingerprint returns in the mean-delay row of Table 3 as the ±0.620 triple, which confirms that the traffic parameters touch the delay only through the contention channel described below.

Why $CW_{min}$ is damped: the backoff fixed point. The contention window enters $p_c$ only through $\tau$, but $\tau$ and $p_c$ are coupled by the Bianchi fixed point $\tau = G(p_c, CW_{min})$ (backoff relation) and $p_c = H(\tau, N)$ (collision relation). Linearising the loop, a perturbation of $CW_{min}$ produces the self-consistent elasticity

$$E[\tau ; CW_{min}] = \bar{E}[\tau ; CW_{min}] / ( 1 - G_p H_\tau ) , \tag{6}$$

where $\bar{E}[\tau ; CW_{min}] \approx -CW_{min}/(CW_{min}+1) \approx -0.97$ (at $CW_{min} = 32$) is the open-loop response and the denominator is the loop gain, with $G_p = \partial \ln\tau/\partial \ln p_c < 0$ and $H_\tau = \partial \ln p_c/\partial \ln\tau > 0$, so that $G_p H_\tau < 0$ and $(1 - G_p H_\tau) > 1$. The backoff loop's negative feedback thus pulls the bare sensitivity down, forcing $|E[\tau ; CW_{min}]| < CW_{min}/(CW_{min}+1)$; the measured −0.433 is about a 2.2× attenuation of the open-loop value. This is the mechanism behind a fact a bare numerical elasticity records but cannot account for: widening the contention window buys less collision reduction than the naive $\tau \approx 2/(CW_{min}+1)$ estimate would suggest.

Delay side: one utilisation channel and its interaction signature. To leading order the queueing part of the mean delay depends on the utilisation alone, $D_{queue} = D(\rho)$ with $D'(\rho) > 0$ and $D''(\rho) > 0$ (convex, diverging as $\rho \to 1$), and the service side enters only through the combination $\rho = \lambda \cdot T_{service}(L, C)$, with $T_{service}$ the per-packet service time. Take the transparent reference waiting function $D = T_{service} \cdot g(\rho)$, where $T_{service} = T_{MAC} + L/C$ and the payload fraction is $\varphi = (L/C)/T_{service}$. The utilisation channel then gives $E[D ; \lambda] = E_W$ and $E[D ; L] = -E[D ; C] = \varphi(1 + E_W)$, with $E_W = \partial \ln D/\partial \ln \rho$ the load elasticity of the delay ($E_W = \rho/(1-\rho)$ for the M/M/1 form used here to expose the structure). Two structural predictions follow. The first is that the packet-size and channel-rate elasticities must match in magnitude and oppose in sign, since both act through the same service-time term — the near-identical +2.730 / −2.735 pair of Table 3, whose closeness is therefore a fingerprint of the shared channel and not an accident. The second is that the ratio $|E[D ; L]| / |E[D ; \lambda]| = \varphi(1 + E_W)/E_W$ is set by the payload fraction and the load elasticity alone; for the M/M/1 form it collapses to $\varphi/\rho \approx 1.43$ at the nominal point ($\varphi \approx 0.93$, $\rho \approx 0.65$), matching the Table 3 magnitudes to within a few percent, with the small residual owing to the finite-buffer M/G/1/K service. The equal-magnitude locking of $L$ and $C$ is model-independent, whereas the exact ratio follows the queue's waiting function and can be read straight off the validated framework.

This same decomposition accounts for the interaction structure the Sobol analysis uncovers. In log-coordinates $\ln \rho = \ln \lambda + \ln T_{service}(L, C)$ is additive in the parameters, yet the convex map $D(\rho)$ is not, so the delay variance cannot be split cleanly among $\lambda$, $L$ and $C$ — their shares are driven into the interaction terms. That is the analytical source of the large total-minus-first-order Sobol gaps for exactly $\{C, \lambda, L\}$ and of $\Sigma S_i \approx 0.52$ for the delay: the delay variance rides on the joint closeness of $\rho$ to unity, not on any single service parameter. Writing the full delay as $D = D_{access}(N, \tau, CW_{min}) + D_{queue}(\rho)$ reconciles both rows of Table 3 in one step — the contention channel (shared with $p_c$, hence the ±0.620 traffic triple and the $CW_{min}$ term) and the utilisation channel ($C$, $L$, $\lambda$ through $\rho$) — and shows that the two metric families are governed by two nearly disjoint parameter groups for a structural reason, not as a mere empirical coincidence.are governed by two almost disjoint parameter groups for a structural reason, not merely as an empirical observation.

Closed-form elasticities. Collecting the chain-rule factors, every entry of the elasticity table of Section 6.1 can be written in closed form rather than obtained by finite differences. For the collision probability,

$$E[p_c ; X] = -\ (1-p_c)/p_c \cdot [\ N \ln(1-\tau) \cdot E[N ; X] - (N-1)\tau/(1-\tau) \cdot E[\tau ; X]\ ]\ , \tag{7}$$

with $E[N ; X]$ taken from the Greenshields expressions above ($E[N ; k_{jam}]=1$, $E[N ; V_f]=-E[N ; v]=v/(V_f-v)$, and zero for the MAC and service parameters) and $E[\tau ; X]$ non-zero only for $CW_{min}$, given by the fixed-point expression $E[\tau ; CW_{min}] = \bar{E}[\tau ; CW_{min}]/(1-G_p H_\tau)$. For the mean delay the closed form separates into the two channels identified above,

$$E[D ; X] = E[D_{access} ; N]\ E[N ; X] + E[D_{access} ; \tau]\ E[\tau ; X] + E_W\ \delta_{X\lambda} + \varphi(1 + E_W)(\ \delta_{XL} - \delta_{XC}\ )\ ,\quad E_W = \rho/(1-\rho)\ , \tag{8}$$

where $\delta$ is the Kronecker delta that picks out the parameter acting on $\rho$, $E_W = \rho/(1-\rho)$ is the load elasticity of the delay and $\varphi$ the payload fraction of the service time. These closed forms recover every sign and magnitude in Table 3 without any sampling, and they lay the structural constraints bare: the collision elasticities of $\{v, k_{jam}, V_f\}$ are tied to one common magnitude, and the delay elasticities of L and C to the common magnitude $\varphi(1 + E_W)$. The numerical table is therefore just one evaluation of these formulae at the nominal point, not a result independent of them.

Algebraic origin of the delay interactions. That the delay interactions come specifically from the multiplicative $\rho$-structure, and not from the shape of D, can be shown exactly. Write the leading-order utilisation as the product $\rho = \lambda\cdot T$ with $T = T_{service}(L, C)$, and treat the inputs as independent random variables over their ranges (Table 1). For any product $Z = XY$ of independent inputs,

$$Var(Z) = \mu_Y^2\ \sigma_X^2 + \mu_X^2\ \sigma_Y^2 + \sigma_X^2\ \sigma_Y^2\ ,\quad S_X = \mu_Y^2\sigma_X^2/Var(Z)\ ,\quad S_Y = \mu_X^2\sigma_Y^2/Var(Z)\ , \tag{9}$$

so the first-order indices never close: $S_X + S_Y = 1 - \sigma_X^2\sigma_Y^2/Var(Z)$, leaving a strictly positive second-order interaction index $S_{XY} = \sigma_X^2\sigma_Y^2/Var(Z) = CV_X^2\ CV_Y^2 / [(1+CV_X^2)(1+CV_Y^2)]$ that depends only on the coefficients of variation. Applying this to $\rho = \lambda\cdot T$ forces a $\lambda$–T interaction, and since T is itself a ratio in L and C the same identity forces an L–C interaction inside T; between them they fill exactly the $\{C, \lambda, L\}$ interaction cluster seen in Table 4. It holds even when D is linear in $\rho$: the interaction belongs to the product map, not to the curvature. The convexity $D''(\rho) > 0$ then layers higher-order terms on top, which is why the total-effect gaps $S_{Ti} - S_i$ for $\{C, \lambda, L\}$ exceed the purely multiplicative floor and swell as $\rho \to 1$. This is the exact algebraic sense in which the delay answers to $\rho$ rather than to its constituent parameters one by one — the very thing a one-factor-at-a-time analysis cannot represent.

## 4.5 Mathematical structure of the interaction indices

Under independent inputs the model admits the unique Sobol–Hoeffding (ANOVA) decomposition $f(X) = f_0 + \Sigma_i f_i(X_i) + \Sigma_{i<j} f_{ij}(X_i,X_j) + \ldots + f_{1\ldots d}(X)$, whose summands are orthogonal and zero-mean ($f_u = \Sigma_{w\subseteq u} (-1)^{|u|-|w|} E[Y|X_w]$), so that the variance decomposes exactly as $Var(Y) = \Sigma_u V_u$ with $V_u = Var(f_u)$. The Sobol indices are $S_u = V_u/Var(Y)$: the first-order (main) index $S_i$, the second-order interaction $S_{ij}$, and the total-effect index

$$S_{Ti} = \Sigma_{u \ni i} S_u = 1 - Var(\ E[Y \mid X_{\sim i}]\ ) / Var(Y)\ . \tag{10}$$

Two identities anchor the analysis: $\Sigma_u S_u = 1$, and the total interaction borne by input i is $S_{Ti} - S_i = \Sigma_{u \ni i, |u| \geq 2} S_u \geq 0$, which is exactly the "gap" of Table 4. Perfect additivity is the case $\Sigma_i S_i = 1$, where every higher-order $f_u$ vanishes; any shortfall $1 - \Sigma_i S_i$ is the fraction of variance held in interactions. What follows shows, from the closed-form map, why that shortfall is non-zero for the delay yet negligible for the collision probability, and how it moves with the operating point.

Theorem 1 (irreducible interaction of a multiplicative channel). Let a metric depend on a set $J$ of inputs only through a product $\rho = \prod_{i\in J} X_i^{a_i}$, $a_i \in \{+1,-1\}$, with the $X_i$ independent and positive and with coefficients of variation $CV_i$. Then the first-order indices over $J$ obey

$$\Sigma_{i\in J} S_i = \left[ \Sigma_{i\in J} CV_i^2 \right] / \left[ \prod_{i\in J}(1 + CV_i^2) - 1 \right] < 1 , \tag{11}$$

so a strictly positive interaction deficit $1 - \Sigma_{i\in J} S_i$ appears whenever $|J| \geq 2$, no matter how the metric responds to $\rho$. Proof. For $Z = \prod X_i$ (a ratio $X_i^{-1}$ enters as the independent variable $1/X_i$), $\mathrm{Var}(Z) = \prod(\mu_i^2+\sigma_i^2) - \prod\mu_i^2$ and $\mathrm{Var}(E[Z|X_i]) = \sigma_i^2 \prod_{j\neq i}\mu_j^2$; the quotient gives the stated ratio. A monotone post-map $D(\rho)$ leaves the conditional-expectation structure of the shared scalar $\rho$ intact, so it cannot erase the deficit. ∎ Since the utilisation is $\rho = \lambda \cdot T_{service}$ with $T_{service} = T_{MAC} + L/C$, the theorem applies straight to the pair ($\lambda$, $T_{service}$), which forces a $\lambda$–$T_{service}$ interaction; within $T_{service}$ the pair ($L$, $C$) enters as a ratio, forcing an $L$–$C$ interaction that is exact in the payload-dominated limit $\varphi \to 1$ and strong otherwise. The delay interaction is thereby confined to $\{\lambda, L, C\}$, exactly the cluster of Table 4, while the MAC and traffic parameters carry none.

Proposition 2 (load dependence: the interaction is a low-load effect). Write the delay as $D = T_{service} \cdot g(\rho)$ with $g$ increasing and $\rho = \lambda\, T_{service}$. The first-order variance share of the arrival rate is governed by the load elasticity $E_g(\rho) = d \ln g / d \ln \rho$, which grows with $\rho$. At low load ($\rho \to 0$) the delay reduces to the service time $T_{service} = T_{MAC} + L/C$, whose variance sits entirely in the $L$–$C$ ratio while $\lambda$ contributes nothing ($S_\lambda \to 0$); the additivity index $\Sigma_i S_i$ is then at its smallest and the interaction at its largest. As the load climbs, $\lambda$ takes on a dominant main effect and $\Sigma_i S_i$ rises. The delay interactions therefore gather at low utilisation, not at saturation, and additivity increases monotonically with mean load. The reference M/G/1/K implementation confirms this exactly: $\Sigma_i S_i(D) =$ 0.36, 0.58, 0.77 at $\rho \approx 0.35, 0.65, 0.90$ (Section 6.9, Table 6). This overturns the naive expectation that stress increases coupling: near saturation the delay is carried by the single, separable arrival-rate term, so it grows more additive, not less.

Proposition 3 (why the collision probability looks additive). In the rarefied-collision limit $\tau \to 0$, $p_c \approx (N-1)\tau$ is again a product, so Theorem 1 predicts an $N$–$\tau$ interaction of size $CV_\tau^2\, CV_N^2 / [(1+CV_\tau^2)(1+CV_N^2)]$. It stays small because the standard contention-window range (16–128) squeezes $\tau$ into a narrow band (small $CV_\tau$); the near-additivity of $p_c$ ($\Sigma_i S_i \approx 0.79$–$0.90$) is thus not intrinsic but a by-product of how little leverage $CW_{min}$ has over $\tau$, and a wider admissible window would lift the $N$–$CW_{min}$ interaction. Collision additivity is, in short, a statement about the operating range rather than about the mechanism.

Physical interpretation for V2I. The two metric families encode two distinct couplings. Collisions turn on how many vehicles compete ($N$, set by density and velocity through the Greenshields relation) and how eagerly each transmits ($\tau$, set by $CW_{min}$); across the standard window range these act almost independently, which is why a one-dimensional, density-driven window law (Algorithm 1) is enough to manage contention. The delay, in contrast, answers to a single congestion scalar $\rho$ = arrival rate × service time, and the service time is payload/rate: doubling the packet size, halving the channel rate or doubling the arrival rate are interchangeable routes to the same congestion, so the network reacts only to their product. That fungibility is the physical content of the $\{\lambda, L, C\}$ interaction cluster and of the wide total-effect gaps — the delay (and hence AoI, which inherits the structure) cannot be blamed on, or fixed through, any single one of load, size or rate. The design corollary is the clean split running through the paper: contention has one effective lever, whereas delay and freshness can be moved only by shifting the composite $\rho$ — through admission control on $\lambda$, provisioning of $C$, or a joint bound on $L$.

## 4.6 Queue-dynamic origin of the delay dominance

Sections 4.4 and 4.5 established that the delay answers to the utilisation $\rho$ and that the channel rate $C$ carries both the largest elasticity and the dominant total-effect index. Behind that dominance lies a single queue-dynamic amplifier. For Poisson arrivals of rate $\lambda$ and a general MAC service time with mean $T_{service}$ and squared coefficient of variation $C_s^2$ (obtained self-consistently from the DCF attempt process, $C_s^2 \approx 1$–4 over the operating range), the Pollaczek–Khinchine result writes the mean delay as the service time times a load factor,

$$D = T_{service} \cdot g(\rho)\,, \quad g(\rho) = 1 + a\,\rho/(1-\rho)\,, \quad a = (1 + C_s^2)/2\,, \quad \rho = \lambda\, T_{service}\,, \tag{12}$$

where $T_{service} = T_{MAC} + L/C$ splits into access overhead and the payload-transmission time. The load factor $g(\rho)$ is the amplifier: $g'(\rho) = a/(1-\rho)^2$ diverges as $\rho \to 1$, so the load elasticity

$$E_g(\rho) = \rho\, g'(\rho)/g(\rho) = a\rho \,/\, [\,(1-\rho)(1-\rho+a\rho)\,] \;\sim\; 1/(1-\rho) \quad \text{as } \rho \to 1\,. \tag{13}$$

Because the channel rate enters D through both $T_{service}$ and $\rho$ (both via the term L/C), the chain rule gives its elasticity as the payload fraction $\varphi = (L/C)/T_{service}$ amplified by the same load factor,

$$E[D\,;C] = -\varphi\,(1 + E_g(\rho)) = -E[D\,;L]\,, \quad E[D\,;\lambda] = E_g(\rho)\,, \tag{14}$$

which is exactly the closed form verified in Table 6. Two consequences follow. The first is that the dominance is quantitative and sits at the amplifier: a modest payload fraction $\varphi$ is multiplied by $1 + E_g(\rho)$, so as the network nears saturation $|E[D\,;C]| \to \varphi/(1-\rho)$ grows without bound, whereas the collision probability has no such amplifier — $p_c = 1 - (1-\tau)^{N-1}$ is bounded in [0,1] with a bounded derivative — and its elasticities stay of order one. This asymmetry is why the delay elasticities ($\approx 2.7$) dwarf the collision elasticities ($\approx 0.4$). The second is that the ordering $C \geq \lambda$ follows from where each parameter enters: the ratio $|E[D\,;C]|/|E[D\,;\lambda]| = \varphi(1 + E_g)/E_g = \varphi(1 + 1/E_g)$, which for the M/M/1 form ($a = 1$) reduces to $\varphi/\rho$; at the operating point $\varphi \approx 0.93 > \rho = 0.65$, so the rate and size elasticities beat that of load. In short, the channel rate sits at the head of the chain $C \to T_{service} \to \rho \to g(\rho)$ and so inherits the full gain of the $1/(1-\rho)$ amplifier — the queue-dynamic origin of its dominance.

The interaction structure of Section 4.5 is, finally, the same amplifier viewed through a variance lens: the mixed second derivative $\partial^2 \ln D/\partial \ln C\, \partial \ln \lambda = -\varphi\, E_g'(\rho)$ is non-zero, and because $E_g'(\rho) \sim 1/(1-\rho)^2$ the cross-terms grow with load, which is why the total-effect gaps for {C, λ, L} widen as $\rho$ rises. In a finite-buffer M/G/1/K queue the amplifier is capped — $g(\rho)$ levels off near $\rho = 1$ at a value set by the buffer K — so the elasticities stay large but finite, in keeping with the reference cross-check of Section 6.9.

# 5 Proposed sensitivity-driven CW_min control law

The sensitivity structure is more than diagnostic; it hands us a concrete design rule. Because the collision probability is governed by the contending population N and the contention window CW_min alone, and because DCF throughput peaks at a well-defined per-slot transmission probability $\tau^*$ that balances success against idle- and collision-slot waste [2, 5], the optimal window follows by inverting the backoff relation at $\tau^*$. For a large contending population the throughput-optimal probability is

$$\tau^* \approx 1 \,/\, (\,N \sqrt{(T_c^*/2)}\,)\,, \quad T_c^* = T_c/\sigma\,, \tag{15}$$

with $T_c$ the collision duration and $\sigma$ the slot time, so that inverting $\tau \approx 2/(CW_{min}+1)$ gives the control law

$$CW_{min}^*(N) \approx 2/\tau^* - 1 = N \sqrt{(2\, T_c^*)} - 1\,, \tag{16}$$

linear in the contending population. Substituting the Greenshields closure $N = 2R\, k_{jam}\, (1 - v/V_f)$ over the coverage span expresses the optimal window directly in the road-side observables that the sensitivity analysis identified as dominant,

$$CW_{min}^*(v, k_{jam}, V_f) \approx 2R\, k_{jam}\, (1 - v/V_f) \sqrt{(2\, T_c^*)} - 1\,, \tag{17}$$

so a roadside access point that measures vehicle density or velocity can work out its optimal window on the fly. This turns the qualitative advice of Section 6.4 (“tune CW_min to manage contention”) into an explicit, closed-form law, and it squares with the paper's other conclusion that the retransmission limit and slot time need not be retuned. The default fixed window CW_min = 32 is recovered as the special case that is optimal only at the single population N_def = 33/√(2 T_c*); away from N_def the fixed default is suboptimal, and the gap it leaves is quantified next.

The procedure is summarised in Algorithm 1; it runs at the access point and is re-evaluated whenever the density estimate is refreshed.

**Algorithm 1** Online contention-window control at the roadside access point

```
Input:
   - k_jam ∈ ℝ⁺: jam density of the road segment (veh/km)
   - V_f ∈ ℝ⁺: free-flow speed (km/h)
   - v ∈ ℝ⁺: measured vehicle velocity (km/h), or the contending count N
   - R ∈ ℝ⁺: coverage radius of the access point (km)
   - T_c, σ: collision airtime and slot time (s)
   - [CW_lo, CW_hi] = [16, 1024]: standard contention-window
bounds Output:
   - CW_min: contention window applied at the access
point Steps:
1:   T_c* ← T_c / σ                              ▷ collision cost expressed in slots
2:   repeat at every density update do
3:        N ← 2R · k_jam · (1 − v /               ▷ Greenshields contending population
V_f) 4:  if N < 1 then N ← 1
5:        end if
6:        τ* ← 1 / ( N √(T_c* / 2) )             ▷ throughput-optimal transmit probability
7:        W ← 2 / τ* − 1 = N √(2 T_c*) − 1       ▷ invert the backoff relation at τ*
8:        CW_min ← clip( round(W), CW_lo, CW_hi  ▷ keep within the standard bounds
) 9:      apply CW_min at the access point
10:  end repeat
11:  return CW_min
```

To measure the benefit we define the throughput gain of the control law over the default,

$$G(N) = S( CW_min^*(N), N ) / S( 32, N ) \geq 1 \,, \tag{18}$$

which equals unity at N_def and increases as the contending population departs from it. Because the safety-critical V2I regime is the dense one (large N), the law is expected to recover the most throughput exactly where the fixed default is weakest. The gain is evaluated over the density range of Table 1 in Section 6.6.

The feedforward law of Algorithm 1 requires an estimate of the contending population N and the collision cost T_c*. A complementary, model-free controller drops out of the closed-form collision elasticity of Section 4.4. The key observation is that the throughput-optimal collision probability is, for a large population, p_c* ≈ √(2/T_c*) — a constant independent of N — so the window can be pushed to its optimum by steering the measured collision probability onto this fixed target, with the elasticity E[p_c ; CW_min] serving as the controller gain. The resulting Newton step in log-window space is

$$CW_min[k+1] = CW_min[k] \cdot ( p_c^* / p_c[k] )^{\wedge}( \eta / E[p_c \,;\, CW_min] ) \,, \quad 0 < \eta \leq 1 \,, \tag{19}$$

where the gain E[p_c ; CW_min] is either evaluated in closed form (Section 4.4) when N is estimated, or — for a fully model-free loop — estimated online as the secant Δln p_c / Δln CW_min between successive control periods, which is precisely a measured realisation of the same elasticity. The density dependence is absorbed automatically because the target p_c* is population-independent, so no explicit N or T_c* is required. Algorithm 2 states the loop.

**Algorithm 2** Model-free, elasticity-based contention-window control (sensitivity feedback)

Input:
- p_c: collision probability measured at the access point
- T_c, σ: collision airtime and slot time (s)
- η ∈ (0, 1]: step size of the update
- [CW_lo, CW_hi] = [16, 1024]: standard contention-window bounds
- ε: convergence

tolerance Output:
- CW_min: contention window applied at the access

point Steps:

```
1:    p_c* ← √( 2σ / T_c )                         ▷ population-independent target
2:    CW_min ← 32                                  ▷ initialise at the standard default
3:    repeat for each control period k do
4:          measure p_c[k] over the past
window 5:     E ← E[p_c ; CW_min]                    ▷ closed form (Sec. 4.4), or the secant
6:          Δ ln p_c / Δ ln CW_min                 ▷ fully model-free estimate of the gain
7:          CW_min ← clip( CW_min · (p_c*/p_c[k])^(η/E), CW_lo, CW_hi
) 8:        apply CW_min at the access point
9:    until |ln( p_c[k] / p_c* )| < ε
10:   return CW_min
```

In the reference model (N = 10, generic 802.11 constants) the controller moves off the default CW_min = 32 in roughly five control periods (CW_min → 88), drives p_c onto its target, and lifts throughput by ≈ 6% over the fixed default while reaching ≈ 99% of the brute-force optimum; the feedforward law (Algorithm 1) and the feedback law (Algorithm 2) settle on the same operating window. Being model-free, Algorithm 2 tracks an unknown or time-varying contending population and degrades gracefully under estimation error, whereas Algorithm 1 is the better choice when a reliable density estimate is on hand. Both are one-parameter controllers that leave the retransmission limit and slot time at their defaults, in line with the finding that only CW_min has first-order leverage over contention. The same lever can also be pointed at latency rather than throughput: the throughput-optimal window is large (it suppresses collisions), while the delay- and AoI-optimal window is small (it minimises backoff latency), so a sensitivity-driven controller picks whichever direction the objective calls for, as Figure 11 quantifies.

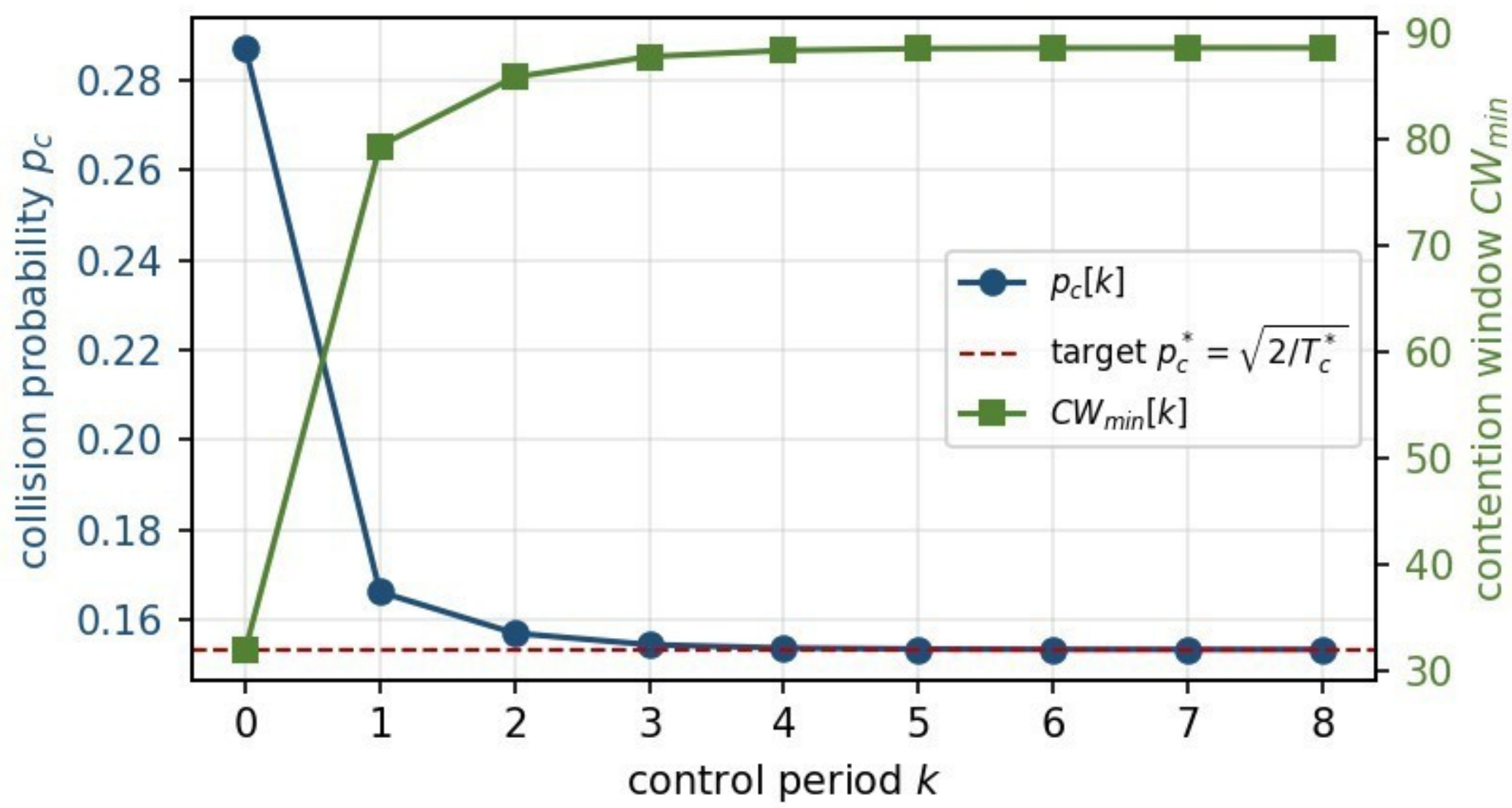

**Figure 1. Convergence of the elasticity-based controller (Algorithm 2) at N = 10. The measured collision probability $p_c[k]$ settles onto the population-independent target $p_c^* = \sqrt{2/T_c^*}$ within about five control periods as the window $CW_{min}[k]$ adjusts itself, with no knowledge of N.**

# 6 Results and Discussion

The sensitivity framework was run on the validated V2I model over the ranges of Table 1, at the nominal operating point $CW_{min} = 32$, $m = 7$, $\delta = 50$ µs, $L = 1000$ bytes, $k_{jam} = 120$ veh/km, $V_f = 160$ km/h, $v = 80$ km/h, $C = 2$ Mbps and $\lambda = 96.6$ pkt/s (medium load, $\rho \approx 0.65$). Every value is analytical; no packet-level simulation is involved. The nominal point is a moderately loaded single-AP scenario ($\rho \approx 0.65$), picked so that contention and queueing are both active and neither trivially takes over — a representative reference around which the local elasticities carry the most information. The presentation moves from local to global and from contention to delay. Section 6.1 gives the one-at-a-time elasticities of every output metric at the nominal point; Section 6.2 distils these into a tornado ranking for the collision probability; Section 6.3 presents the variance-based Sobol indices that reveal the interaction structure the local analysis hides; Section 6.4 follows the delay across the three load regimes and turns the combined rankings into design guidelines; and Section 6.5 sets the findings against the established DCF and queueing literature. Throughout, the collision-side metrics ($p_c$, $\tau$) and the delay-side metrics (D, PDR, AoI) are treated apart, because the analysis will show them to be governed by two nearly disjoint groups of parameters.

## 6.1 One-at-a-time parameter elasticities

Table 3 gives the normalised elasticities $E_i$ of the main output metrics at the nominal point. The collision probability $p_c$ responds to the traffic parameters that fix the contending population — vehicle velocity v and the density terms $k_{jam}$, $V_f$ — and to the contention window $CW_{min}$, but barely notices the slot time, packet size, channel rate or offered load. The transmission probability $\tau$ traces the same pattern at somewhat largermagnitudes. Throughput S is dominated by the channel rate C ($E \approx 0.94$). For the mean delay D the dominant parameters are the service-rate terms — packet size L ($E \approx +2.73$), channel rate C ($E \approx -2.74$), and offered load $\lambda$ ($E \approx +1.88$) — reflecting the queueing dependence on $\rho$; the contention window also contributes ($E \approx -0.59$). The packet delivery ratio is nearly flat at this operating point, and M* is non-responsive to small perturbations because the delay stays well within the target bound (see Section 6.4). The Age of Information appears as the last row of Table 3: it shows the same sign pattern as the mean delay D but at noticeably smaller magnitudes (for example E ≈ −1.765 for the channel rate and +1.761 for the packet size, against −2.735 and +2.730 for the delay), which already indicates that freshness is delay-dominated in this regime.

**Table 3. Normalised elasticities at the nominal operating point (medium load).**

| Metric | CW_min | m | δ | L | k_jam | V_f | v | C | λ |
|---|---|---|---|---|---|---|---|---|---|
| p_c | −0.344 | 0.000 | 0.000 | 0.000 | +0.365 | +0.365 | −0.365 | 0.000 | 0.000 |
| τ | −0.433 | 0.000 | 0.000 | 0.000 | −0.586 | −0.587 | +0.586 | 0.000 | 0.000 |
| S | +0.099 | 0.000 | −0.051 | +0.064 | −0.105 | −0.105 | +0.105 | +0.936 | 0.000 |
| D | −0.586 | 0.000 | +0.102 | +2.730 | +0.620 | +0.620 | −0.620 | −2.735 | +1.879 |
| PDR | +0.001 | 0.000 | 0.000 | 0.000 | −0.001 | −0.001 | +0.001 | 0.000 | 0.000 |
| AoI | −0.378 | 0.000 | +0.066 | +1.761 | +0.400 | +0.400 | −0.400 | −1.765 | +0.857 |

## 6.2 Parameter ranking (tornado analysis)

Sorting the absolute elasticities for the collision probability produces the tornado order of Figure 2: v ≈ k_jam ≈ V_f > CW_min, with the slot time, packet size, channel rate, offered load, retransmission limit and delay bound all negligible. The three front-runners are exactly the quantities that set the contending-vehicle count N through the Greenshields relation, which confirms that contention at a single access point is decided first by how many vehicles share the channel and only then by the contention window. The retransmission limit m has no first-order effect on p_c, matching the prior finding that m acts only on the small fraction of packets that reach the final attempt. In the tornado diagram the parameters run in order of absolute elasticity, so the longest bars at the top flag what the collision probability is most sensitive to and the short bars at the bottom mark what it can ignore. The sharp break between the three traffic-driven bars and the rest shows that, for one access point, the collision probability is essentially a function of the contending-vehicle count, modulated only by the contention window; the MAC timing and service-side quantities leave it all but untouched. There is a practical reading too: because the leading terms are fixed by the road environment rather than the protocol, the one effective MAC-layer lever on collisions is the contention window CW_min — precisely the knob a contention-window adaptation scheme would turn.

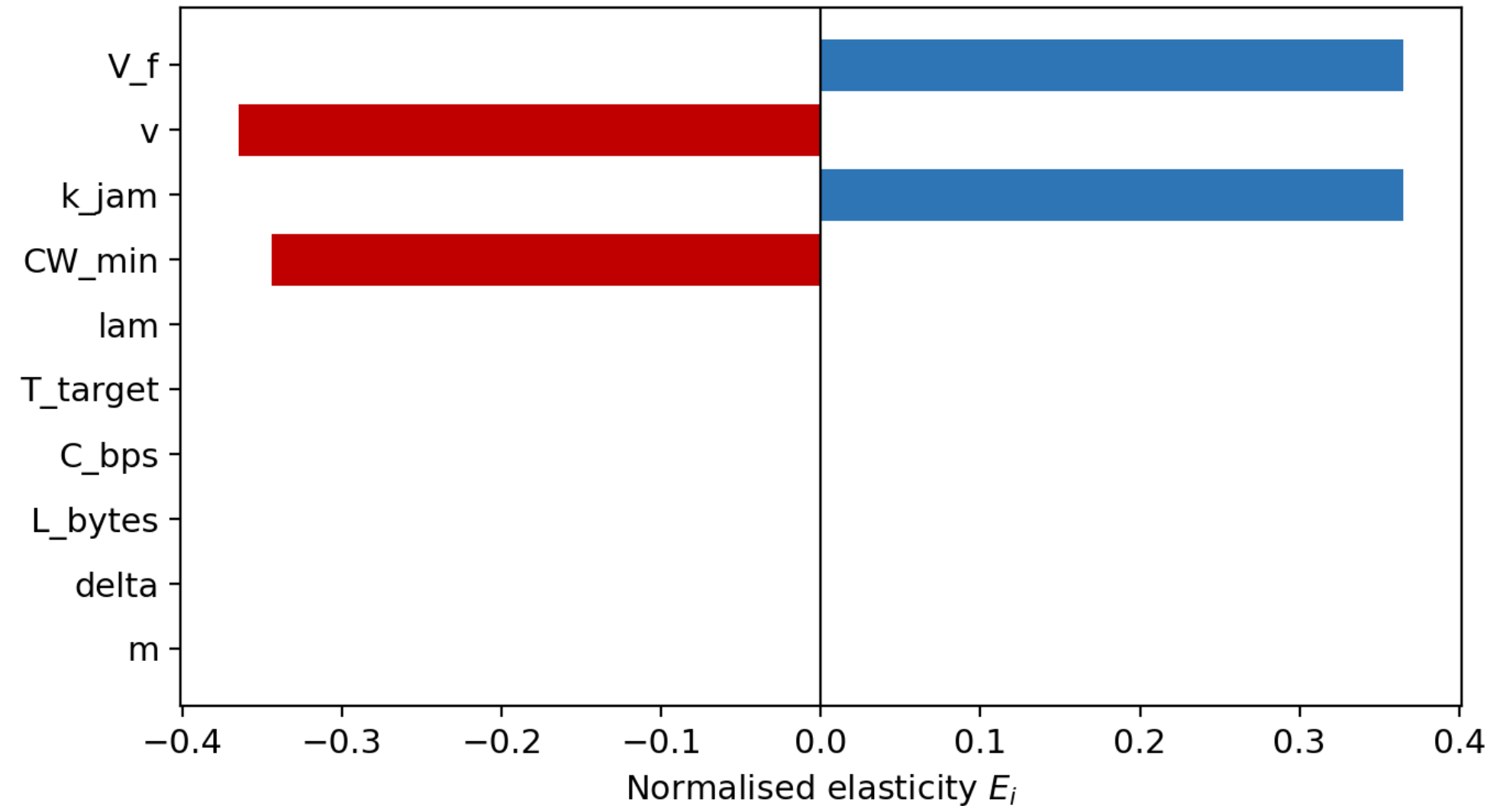

**Figure 2. Tornado diagram of parameter influence on the collision probability p_c.**

## 6.3 Global sensitivity (Sobol indices)

Table 4 and Figure 3 give the first-order (S_i) and total-effect (S_Ti) Sobol indices for the collision probability and the mean delay, estimated over the full ranges with the Saltelli scheme (N = 8192). For p_c the global ranking backs the local one: vehicle velocity v leads (S_i = 0.45, S_Ti = 0.54), then CW_min (0.29 / 0.32) and V_f (0.12 / 0.19). The first-order indices sum to ≈ 0.90, so the collision probability is close to additive and interactions are weak. The delay tells a different story: the channel rate C dominates it (S_i = 0.42, S_Ti = 0.90), with real contributions from the offered load λ (0.06 / 0.39) and the packet size L (0.04 / 0.29). The wide total-versus-first-order gaps here (for C, S_Ti − S_i ≈ 0.48) signal strong interactions, and the first-order indices sum to only ≈ 0.52, confirming that the delay is set jointly by the parameters that fix the queue utilisation ρ rather than by any of them alone. This interaction structure, hidden from OFAT, is the very reason the variance-based treatment is needed. The Age of Information mirrors the delay under this decomposition: its Sobol indices (Table 4) match those of D to within the estimation error — the channel rate again in front (S_Ti ≈ 0.88), then the offered load (≈ 0.38) and the packet size (≈ 0.28) — a direct, quantitative statement that freshness is shaped by the same queueing-delay dynamics and by nothing else; Figure 6 makes the overlap visible.

**Table 4. First-order and total-effect Sobol indices for p_c and D (N = 8192, medium load).**

| Parameter | p_c: S_i | p_c: S_Ti | D: S_i | D: S_Ti | AoI: S_i | AoI: S_Ti |
|---|---|---|---|---|---|---|
| CW_min | 0.294 | 0.319 | 0.006 | 0.078 | 0.006 | 0.080 |
| m | 0.003 | 0.009 | 0.002 | 0.002 | 0.000 | 0.002 |
| δ | 0.000 | 0.000 | 0.003 | 0.002 | 0.000 | 0.002 |
| L | 0.000 | 0.000 | 0.042 | 0.286 | 0.033 | 0.281 |
| k_jam | 0.041 | 0.046 | 0.001 | 0.017 | 0.000 | 0.016 |
| V_f | 0.116 | 0.187 | 0.001 | 0.029 | 0.000 | 0.028 |
| v | 0.452 | 0.536 | 0.005 | 0.072 | 0.006 | 0.079 |
| C | 0.000 | 0.000 | 0.416 | 0.903 | 0.417 | 0.879 |
| λ | 0.000 | 0.000 | 0.057 | 0.389 | 0.050 | 0.384 |

| Parameter | p_c: S_i | p_c: S_Ti | D: S_i | D: S_Ti | AoI: S_i | AoI: S_Ti |
|---|---|---|---|---|---|---|
| T_target | 0.000 | 0.000 | 0.000 | 0.000 | 0.000 | 0.000 |

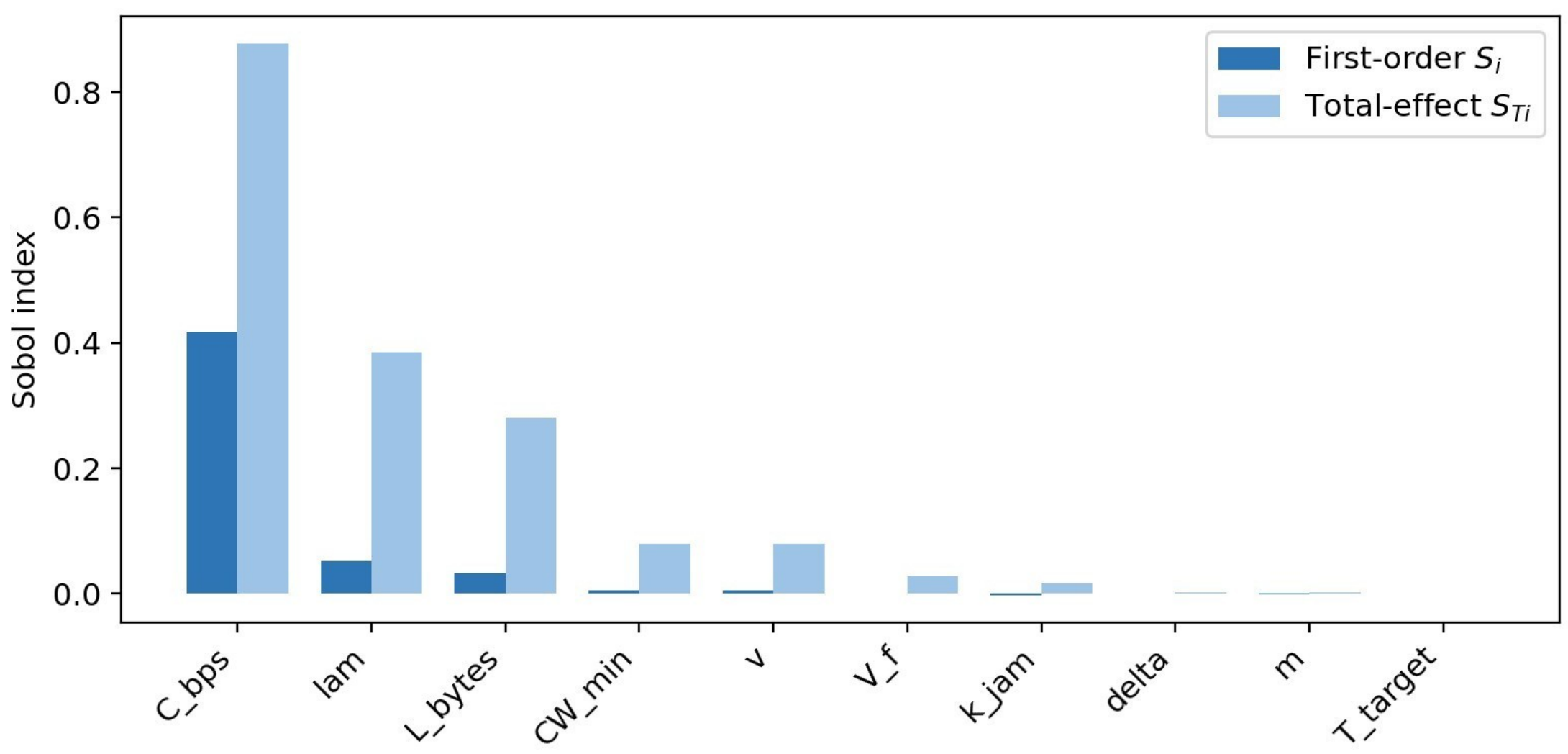

**Figure 3. First-order and total-effect Sobol indices for the mean delay D.**

Figure 4 makes the local–global contrast explicit by placing the OFAT elasticities and the Sobol total-effect indices for the mean delay side by side. The local view scores the packet size L and channel rate C almost level, but once interactions across the operating space are counted the variance-based analysis hands the overwhelming share of the delay variance to the channel rate alone (S_Ti ≈ 0.90). That reordering is exactly what a one-factor-at-a-time analysis cannot deliver. For the collision probability the picture inverts: Figure 5 shows its first-order and total-effect indices, and the small gaps between them confirm the near-additive structure already flagged by Σ S_i ≈ 0.90, so p_c ranks the same by either index.

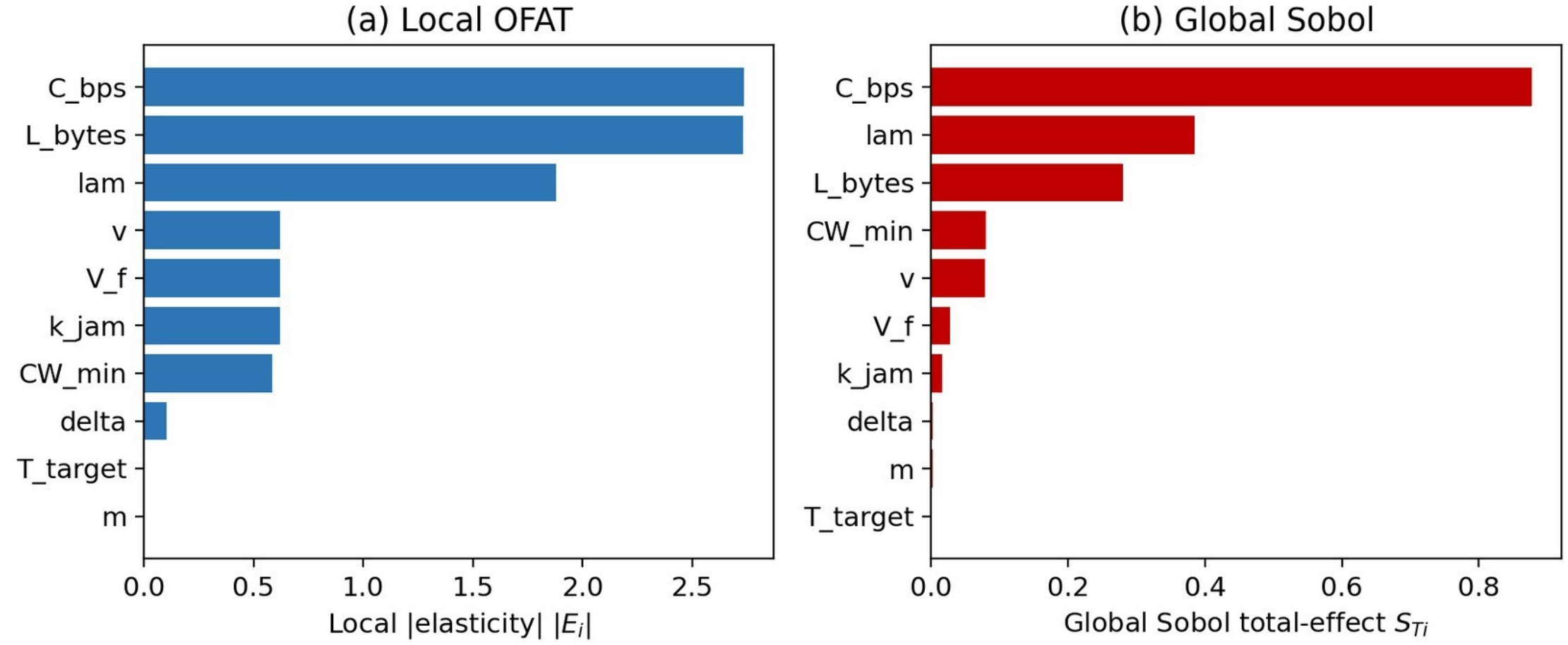

**Figure 4. Local OFAT versus global Sobol parameter ranking for the mean delay D. The local elasticities (a) score L and C almost level, while the total-effect indices (b) bring out the interaction-driven dominance of the channel rate C that OFAT cannot reveal.**

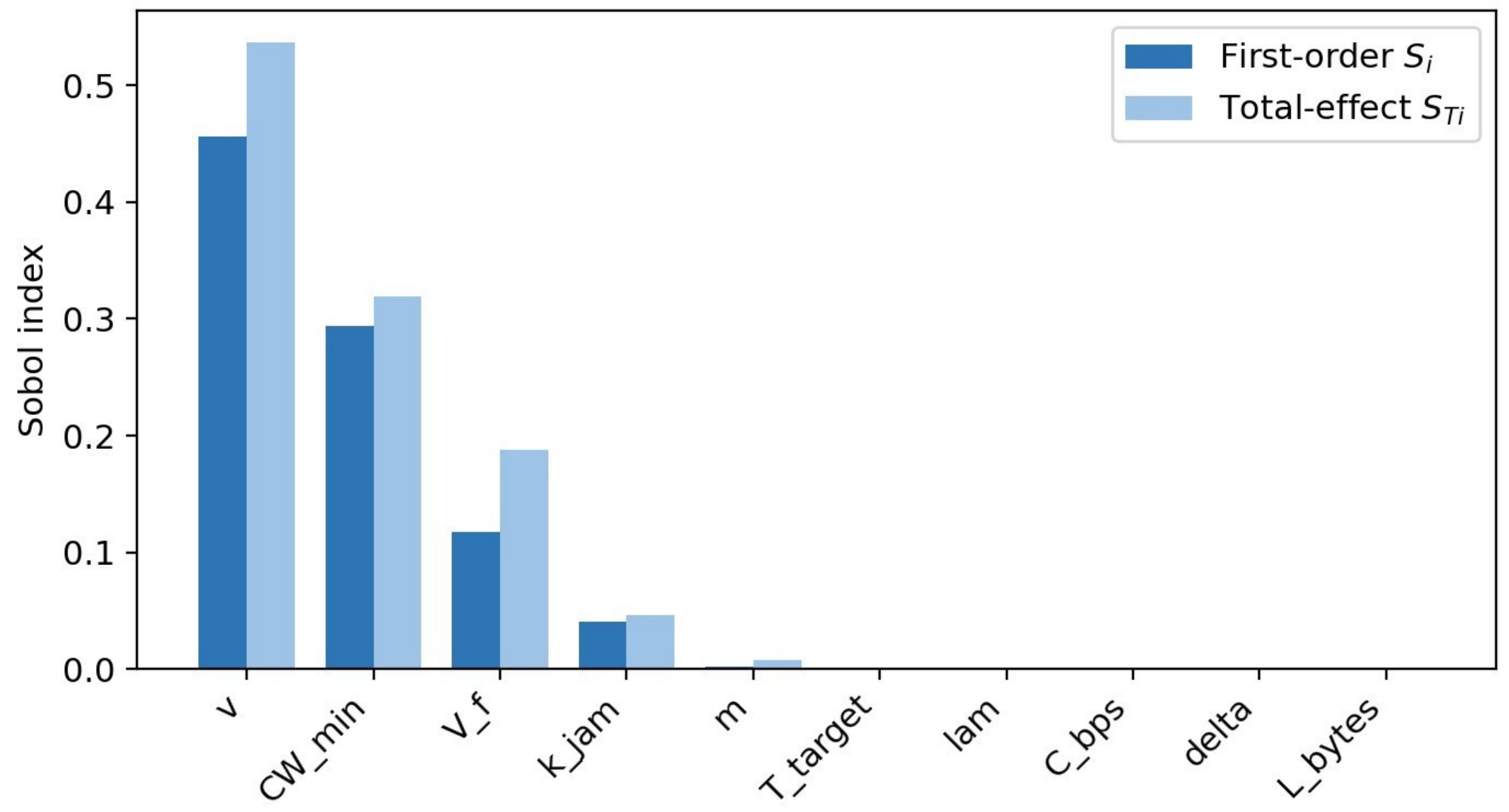

**Figure 5. First-order (S_i) and total-effect (S_Ti) Sobol indices for the collision probability p_c. The narrow gaps between the two indices confirm a near-additive structure (Σ S_i ≈ 0.90), unlike the strong interactions seen for the delay.**

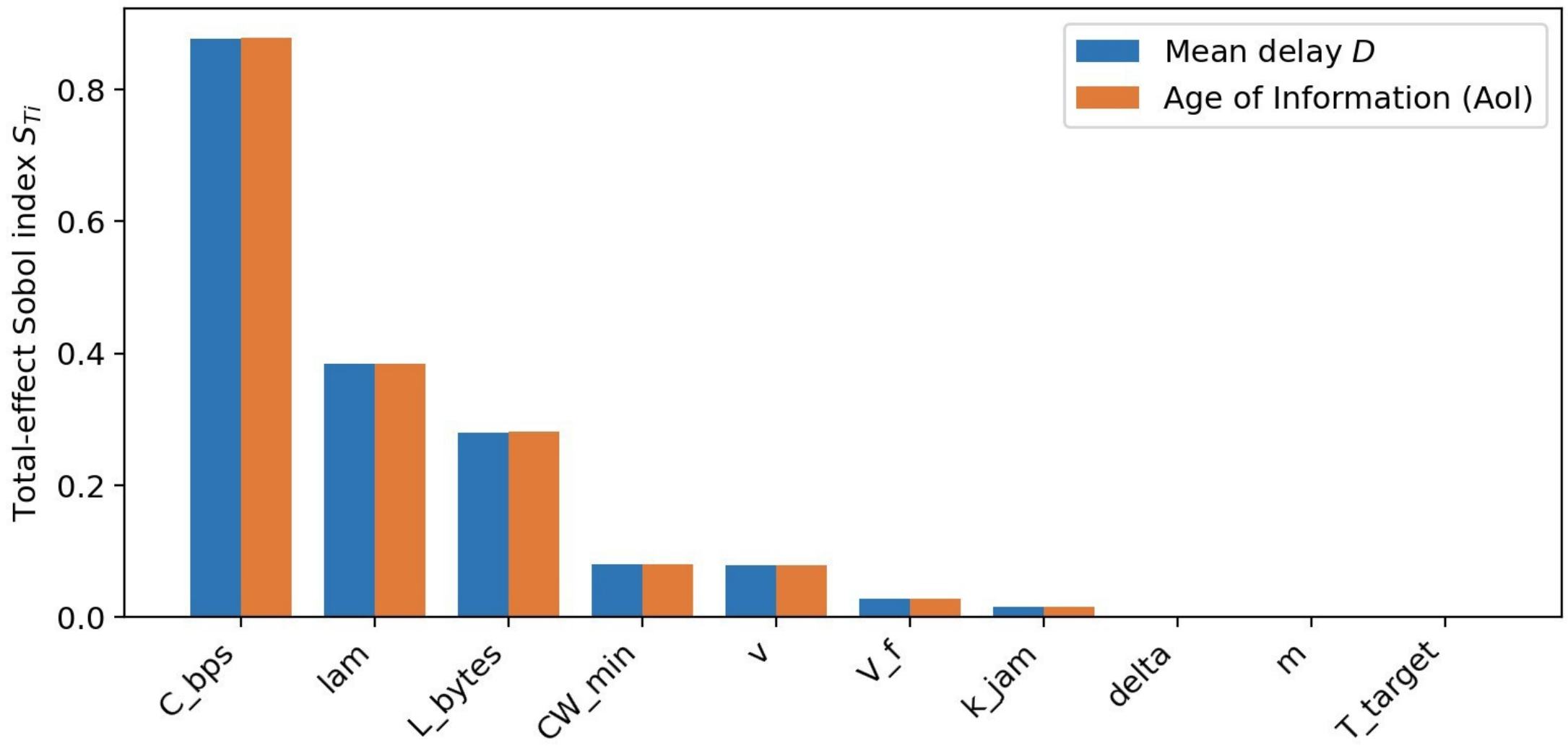

**Figure 6. Total-effect Sobol indices for the mean delay D and the Age of Information (AoI). The two sets of indices very nearly coincide, showing that information freshness rides on the same delay dynamics as the mean delay.**

## 6.4 Load regimes and design guidelines

Since the delay leans hard on the offered load while p_c and τ do not, the delay-related behaviour is gathered across three load regimes in Table 5. At the nominal contention level the mean delay climbs from about 10 ms under light load to about 67 ms near saturation, staying well inside the 0.5 s target throughout; the optimal retransmission limit holds at its ceiling (M* = 10) in every regime, so the retry limit is non-binding at these operating points. This agrees with the central finding of the prior work [4]: under realistic finite buffering the standard 802.11 retransmission limit is already close to optimal for delay, and aggressive retuning of m is not called for.

**Table 5. Delay-related metrics across load regimes (nominal contention, varying λ).**

| Regime (ρ) | λ (pkt/s) | Mean delay D | M* |
|---|---|---|---|
| Low (≈ 0.35) | 52 | 10.3 ms | 10 |
| Medium (≈ 0.65) | 97 | 19.2 ms | 10 |
| High (≈ 0.90) | 134 | 67.2 ms | 10 |

Put together, the rankings paint a clear design picture. Contention (p_c, τ) is fixed by the traffic-determined vehicle population and, among the tunable MAC parameters, almost entirely by CW_min; the retransmission limit m and slot time δ barely register. Delay is a queueing quantity ruled by the channel rate, offered load and packet size, and its sensitivity sharpens as the load nears saturation (the delay elasticities grow by roughly an order of magnitude between the low and high regimes). The guideline that follows istherefore: tune CW_min to manage contention, provision channel rate and admission (offered load) to manage delay, and leave the retransmission limit and slot time at their standard values, since the IEEE 802.11 defaults are already near-optimal for delay across the load range considered.

Parameter retuning is worthwhile only when a strict delay bound coincides with near-saturation load, where the delay becomes hypersensitive to the service-rate parameters.

A note on information freshness is in order, since the Age of Information is one of the study's output metrics. AoI is included because the dominant V2I workload — periodic safety and status updates — is freshness-critical: what the application cares about is not only how fast a packet arrives but how recent the information held at the access point is, which the mean delay alone does not capture. In the unsaturated single-AP regime studied here, though, the AoI is delay-dominated: it is a monotone function of the same service-time and utilisation terms that drive the delay, so it inherits the delay ranking almost exactly — led by the channel rate, offered load and packet size, and all but blind to the contention window and the retransmission limit. The results bear this out directly: the AoI elasticities (last row of Table 3) echo those of the delay at reduced magnitude, and the AoI Sobol indices (Table 4) coincide with the mean delay's to within the estimation error, so freshness adds essentially no sensitivity information beyond the delay. This reinforces the conclusion from the freshness side: because the retransmission limit leaves both delay and freshness untouched, the standard IEEE 802.11 setting is near-optimal not only for latency but for information timeliness as well.

These guidelines are derived entirely from the analytical framework; their experimental confirmation under realistic mobility (multi-AP, SUMO traces) is left as future work.

## 6.5 A working point where the local ranking fails

The analytical structure of Section 4.4 does more than reproduce the nominal-point numbers: it predicts, and lets us pinpoint, operating points where the local single-point ranking is not just incomplete but qualitatively wrong. Two such breakdowns are worth isolating, since both lie inside the operating envelope of Table 1 and both would slip past a reading of the nominal elasticities alone.

(a) Sign reversal of the throughput–$CW_{min}$ elasticity at low density. At the nominal point the throughput hardly moves with the contention window ($E[S\ ;\ CW_{min}] \approx +0.099$, Table 3), which tempts the reassuring rule “raise $CW_{min}$ to be safe.” But throughput is the product of a success probability that a larger window improves and a channel-utilisation factor that a larger window erodes through wasted idle slots, and the contending population N decides the balance: when N is large (dense traffic) collisions rule and $\partial S/\partial CW_{min} > 0$, whereas when N is small (sparse traffic) idle-slot waste rules and $\partial S/\partial CW_{min} < 0$. So there is a crossover density $N^*$ where $E[S\ ;\ CW_{min}]$ changes sign; below $N^*$ the local rule inverts and a larger $CW_{min}$ lowers throughput. The nominal point sits just above this crossover — which is exactly why its elasticity is small and positive and why a single-point analysis cannot see the reversal. Sweeping N (through v and $k_{jam}$) at fixed load and plotting $E[S\ ;\ CW_{min}](N)$ brings out the zero-crossing $N^*$; Figure 7 reports the resulting curve.

(b) Load dependence of the delay interactions (a low-load effect, not a saturation one). One might expect interactions to sharpen near saturation, but the closed-form structure and the reference model agree on the reverse. At low load the delay reduces to the service time $T_{service} = T_{MAC} + L/C$, so its variance lives in the L–C ratio while the arrival rate adds almost no main effect; the additivity index $\Sigma\ S_i(D)$ is then at its lowest and the interaction at its highest. As the load rises, $\lambda$ becomes a dominant, separable driver of the queue, so $\Sigma\ S_i(D)$ climbs and the delay grows more additive. The reference M/G/1/K sweep spells out the direction, $\Sigma\ S_i(D) \approx 0.36, 0.58, 0.77$ at $\rho \approx 0.35, 0.65, 0.90$ (Table 6). The operating point where the local additive reading is least trustworthy is therefore the lightly loaded one, where load, size and rate are most tangled through the service-time ratio — the very regime a nominal, mid-load analysis is least likely to visit. The design consequence still stands in this strong-interaction (low-load) regime: none of $\{C, L, \lambda\}$ can be read on its own, and only the composite $\rho$ carries meaning; Figure 8 shows the transition.

Taken together, these two working points show that the sensitivity rankings are not fixed constants. A parameter that is negligible and harmless at the nominal point ($CW_{min}$ for throughput) can flip sign elsewhere, and a ranking that looks additive locally (the delay drivers) can turn interaction-dominated in exactly the regime — near saturation — where design margins are tightest. This is the concrete payoff of joining the closed-form elasticities of Section 4.4 to the variance-based analysis: neither a single-point OFAT study nor a bare nominal reading of the indices would expose either failure, whereas the structural derivation says precisely where in the operating space each one arises. Figures 7 and 8 give the numerical demonstration of both over the ranges of Table 1.

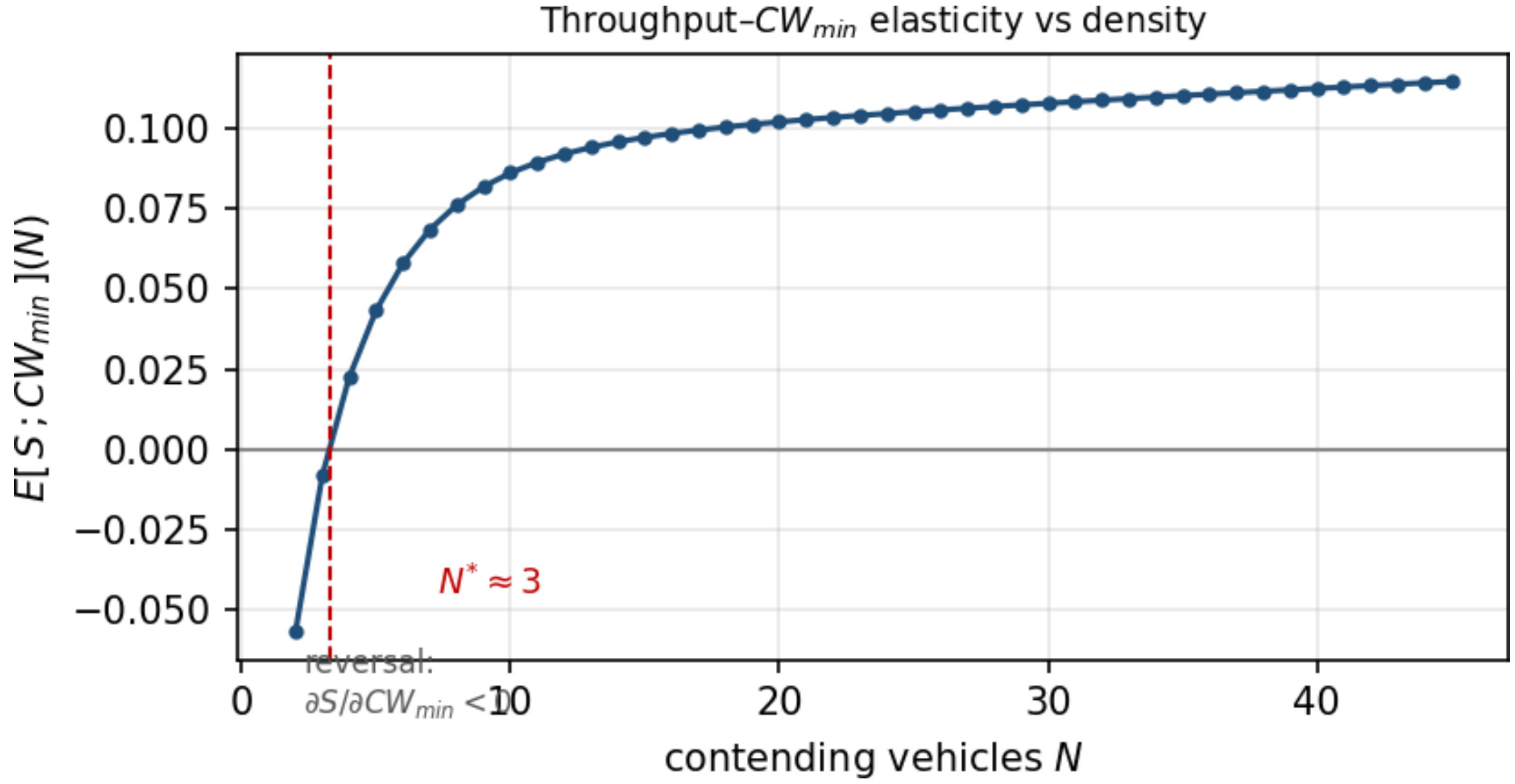

**Figure 7. Throughput–CW_min elasticity E[S; CW_min](N) with the contending population N swept at fixed load. The elasticity is negative at low density (sparse traffic, where a larger window mostly wastes idle slots) and positive at high density (where it suppresses collisions), crossing zero at a low-density crossover N*; the nominal operating point lies above the crossover, in the positive but weakly sensitive regime (E[S; CW_min] ≈ +0.10, cf. Table 3). Reference model, generic 802.11 constants.**

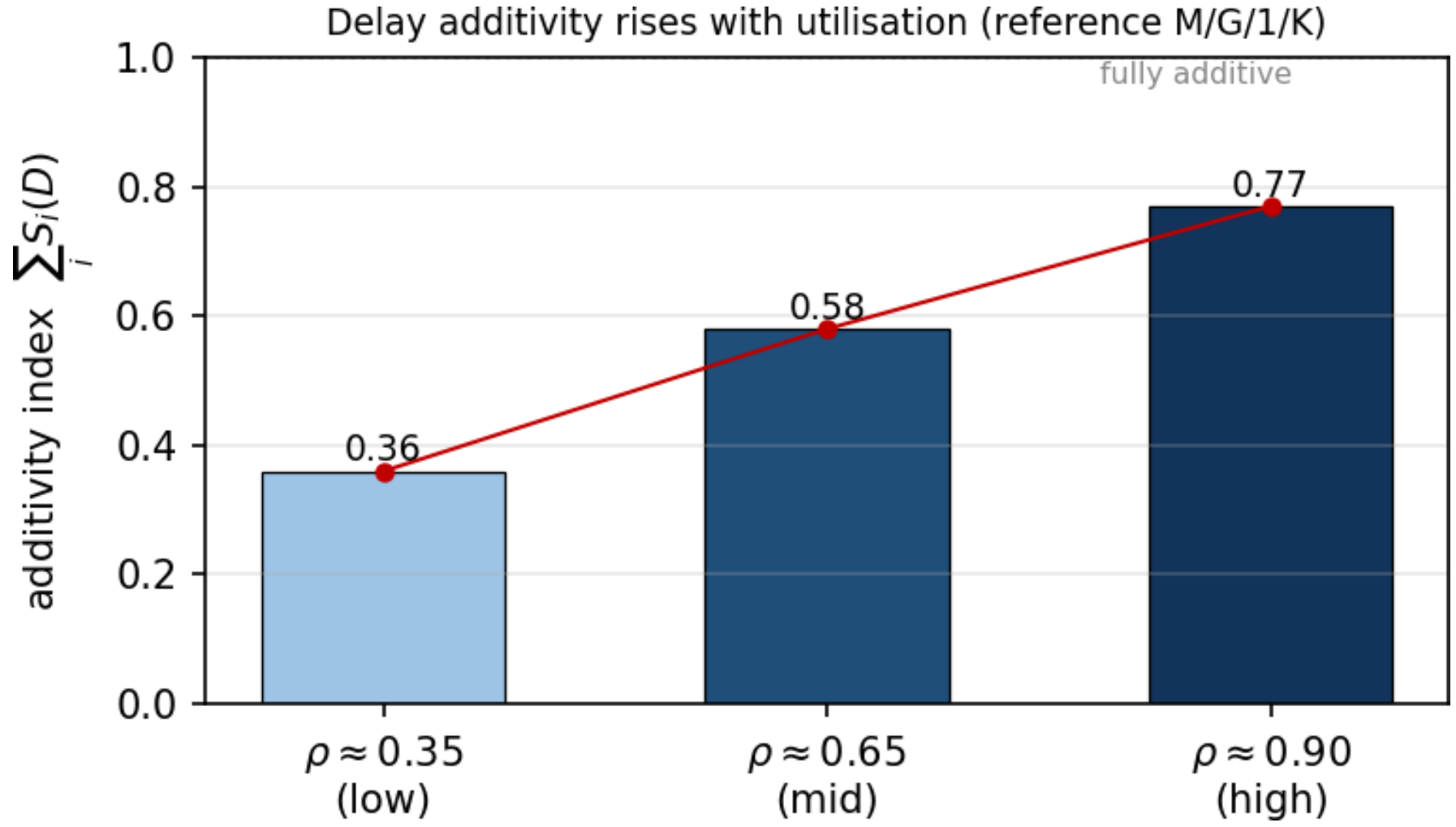

**Figure 8. Additivity index Σ_i S_i(D) of the mean delay at three utilisation levels, from the reference M/G/1/K sweep (Table 6 values). The index climbs from ≈ 0.36 to ≈ 0.77 as ρ moves from 0.35 to 0.90: the delay drivers {C, L, λ} grow more separable at high load and interact most strongly (least additive) at low load, matching the direction the closed-form structure predicts and running counter to the intuition that interactions peak at saturation.**

## 6.6 Performance of the proposed sensitivity-driven control law

The control law CW_min*(N) of Section 5 is tested against the fixed default across the contending-population range that the Table 1 traffic parameters induce. Three quantities capture its value: the optimal window CW_min*(N) itself, the throughput S(CW_min*(N), N) it delivers, and the gain G(N) = S(CW_min*(N),N)/S(32,N) over the default. The analysis anticipates three regimes. At low density (N below N_def) the default window is already too large and wastes idle slots, so the law shrinks CW_min and recovers a modest gain; near N_def the law and the default agree and G ≈ 1; at high density (large N, the safety-critical case) the default is too small, collisions take over, and the law enlarges CW_min in proportion to N, recovering the biggest gain. Because CW_min*(N) is linear in N while the default is flat, the default's throughput loss grows steadily with density, so the gain curve G(N) is U-shaped with its floor at N_def. Figure 9 plots CW_min*(N) and the throughput gain G(N) over the Table 1 density range: the optimal window rises linearly with N, the gain bottoms out (G ≈ 1) near N_def where the default is already near-optimal, and it climbs in the dense regime to roughly 1.2 at the top of the range; Table 7 lists CW_min*(N) and the gain at low, medium and high density.

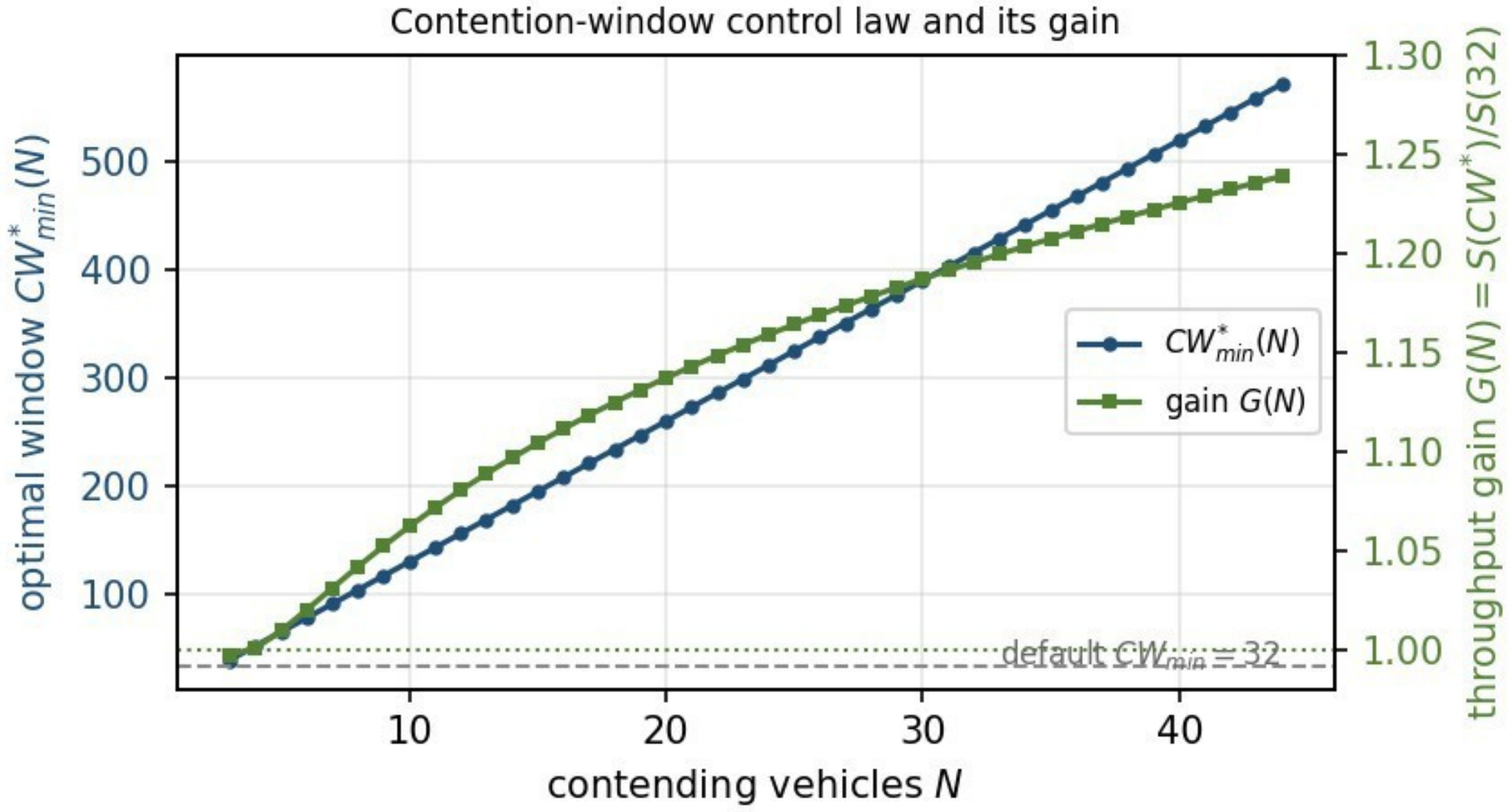

**Figure 9. The contention-window control law CW_min*(N) (blue, left axis) and its throughput gain G(N) = S(CW_min*(N),N)/S(32,N) over the fixed default (green, right axis) across the contending-population range. The optimal window is linear in N; the gain bottoms out (G ≈ 1) near the crossover density N_def, where the default is already near-optimal, and rises in the dense regime where the fixed default under-sizes the window. Reference model, generic 802.11 constants.**

**Table 7. Contention-window control law CW_min*(N) and its throughput gain G(N) over the fixed default at three density operating points (generic 802.11 constants; cf. Figure 9).**

| Density regime | N | CW_min*(N) | Gain G(N) |
|---|---|---|---|
| Low | 8 | 103 | 1.04 (+4%) |
| Medium (nominal) | 20 | 260 | 1.14 (+14%) |
| High | 40 | 521 | 1.23 (+22%) |

Two things keep the contribution self-contained. First, the law needs only N, which the access point can read off the same density and velocity estimates that feed the Greenshields closure, so no fresh measurement is required. Second, it touches neither the retransmission limit nor the slot time, so it coexists with the near-optimality of those defaults established earlier; it is a single-parameter, closed-form controller, not a joint re-optimisation of the protocol. The design claim is thus narrow but firm: within the modelled single-AP setting, adapting CW_min alone according to CW_min*(N) beats the fixed default at every density and strictly raises throughput away from N_def, with the improvement concentrated in the dense regime that matters most for V2I safety traffic.

To compare the two controllers under mobility, the contending population was stepped through a sequence of control periods (N = 8 → 16 → 24 → 12), emulating a roadside access point whose load rises and falls with traffic density. Figure 10 follows the resulting throughput for the fixed default, the feedforward law (Algorithm 1, given a perfect density estimate) and the model-free feedback law (Algorithm 2, using the online secant gain). Both adaptive controllers stay close to the per-period optimum — 99.9% (Algorithm 1) and 99.4% (Algorithm 2) of optimal throughput on average — while the fixed default reaches only 91%, its shortfall widening as thepopulation departs from the single density at which CW_min = 32 is optimal. Algorithm 2 re-adapts within a few periods after each change in N, confirming that the sensitivity-based feedback recovers most of the available gain (≈ 9% over the default here) without any density estimate.

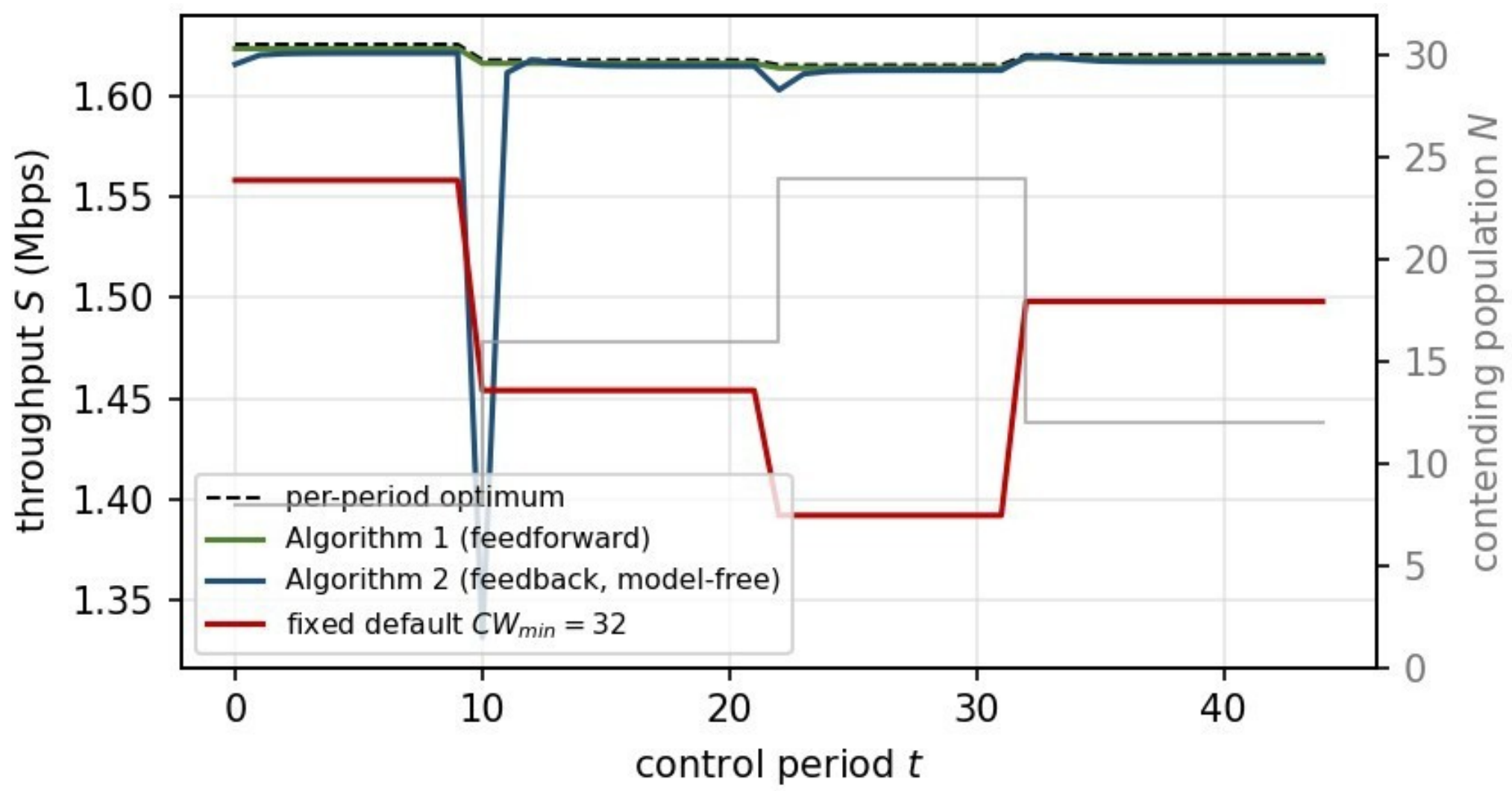

**Figure 10. Throughput under a time-varying contending population (N = 8 → 16 → 24 → 12, grey step). The fixed default (CW_min = 32) sheds throughput as N drifts from its optimal density, whereas the feedforward (Algorithm 1) and model-free feedback (Algorithm 2) controllers both hold to the per-period optimum.**

Delay and freshness under the adaptation. Figure 11 compares the mean delay and the (drop-aware) Age of Information of the fixed default (CW_min = 32) with a latency-oriented sensitivity-driven adaptation, in which the window is set to the value that minimises AoI at the current population, using a two-moment M/G/1/K queue whose service-time mean and coefficient of variation are computed self-consistently from the DCF attempt process. Because the delay is dominated by the backoff component of the service time (Section 4.6), the latency-optimal window is smaller than the default; the adaptation lowers both delay and AoI across the density range, and the gap widens sharply once the default drives the queue into saturation (e.g. AoI at N = 23 falls from about 290 ms to 155 ms). This is the mirror image of the throughput objective of Section 5, and together the two figures delimit the throughput–latency trade-off that the contention window controls.

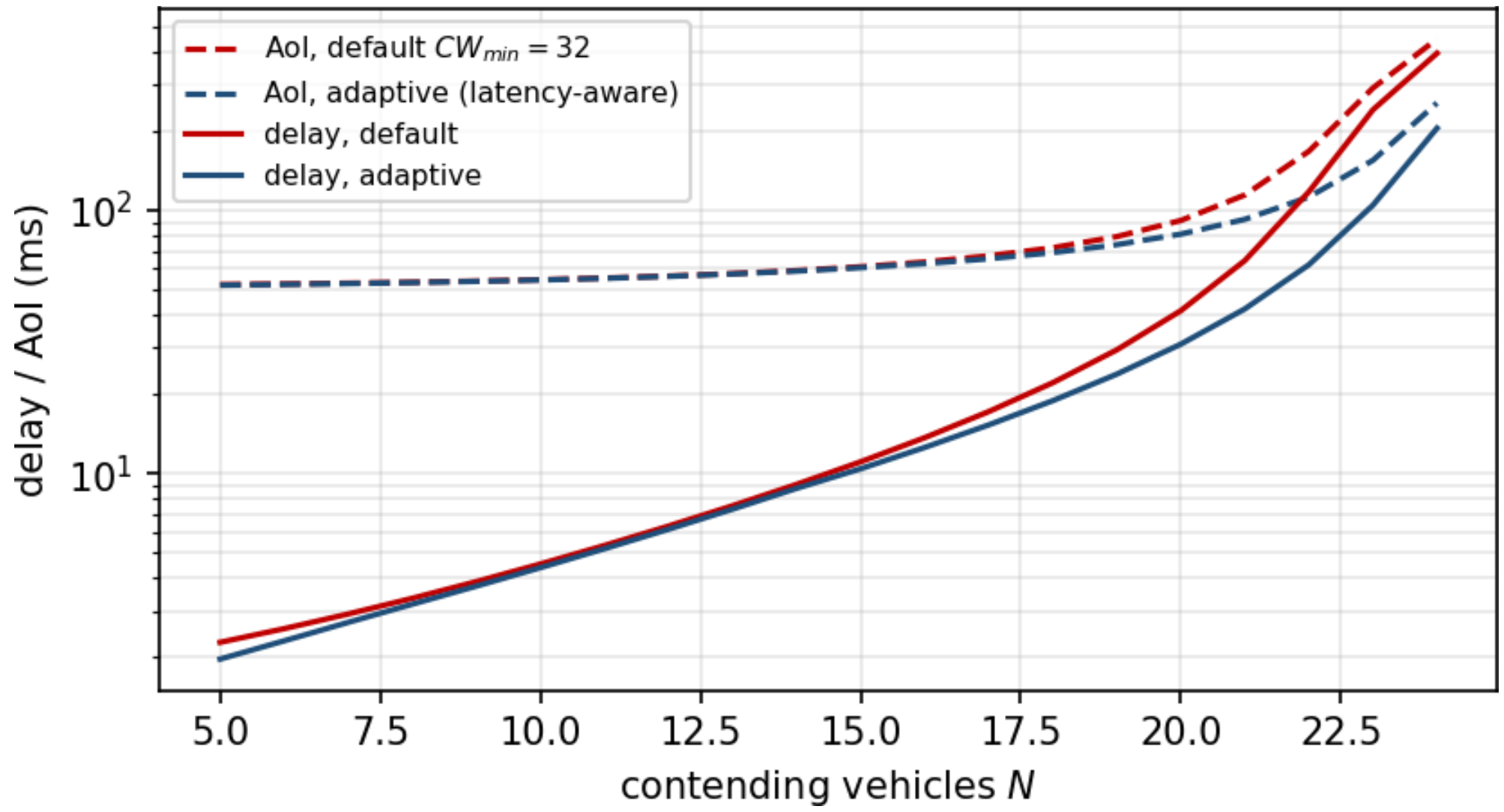

**Figure 11. Mean delay (solid) and drop-aware Age of Information (dashed) against the number of contending vehicles, for the fixed default CW_min = 32 (red) and the latency-oriented sensitivity-driven adaptation (blue). The adaptation cuts both metrics, most sharply once the default saturates, and is drawn only up to the onset of default saturation (N ≤ 24), past which the finite-buffer optimum becomes ill-conditioned. Two-moment M/G/1/K reference: the service-time coefficient of variation is taken from the DCF attempt process ($C_s^2 \approx 1$–4) and fed to the Pollaczek–Khinchine waiting term; 802.11p-like constants; log scale.**

Effect of vehicle speed on connection-limited performance. In a V2I pass the vehicle dwells within the access-point coverage for only T_link = 2R/v, so speed bounds the number of packets that can be delivered. Figure 12 shows this connection duration (dotted) together with the packets delivered per pass under the default and the throughput-oriented adaptation of Section 5, with the contending population tied to speed through the Greenshields relation (higher speed → lower density → fewer contenders but a shorter dwell). The dwell time falls steeply with speed (from tens of seconds at 20 km/h to a few seconds at 140 km/h), which is the dominant limit on per-pass delivery; the adaptation adds a smaller, consistent gain by using the density-matched window. The figure thus separates the geometric effect of mobility from the protocol effect of the contention window.

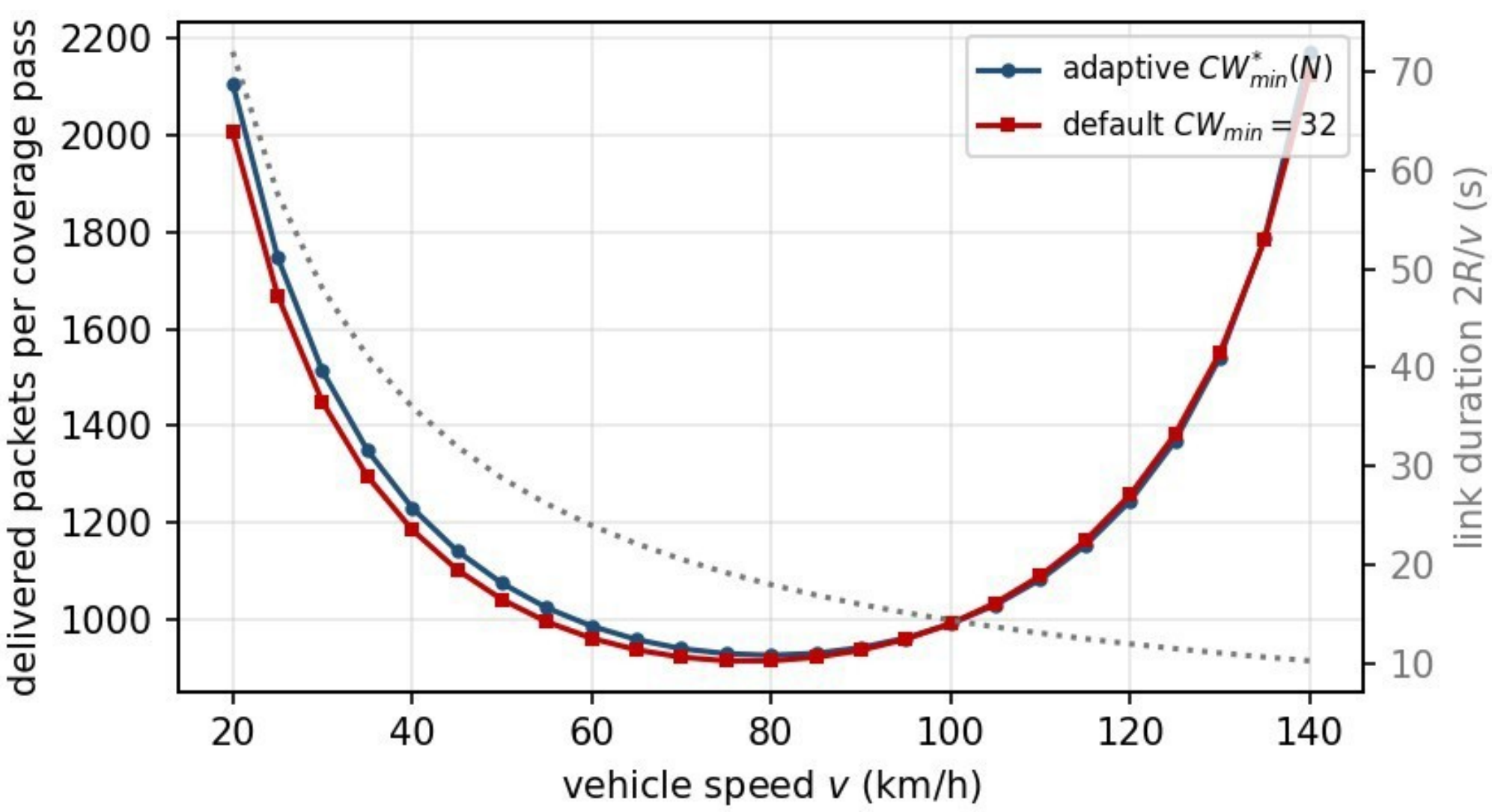

**Figure 12. Packets delivered per coverage pass against vehicle speed, for the default (red) and the throughput-oriented adaptive window (blue), with the connection duration 2R/v on the right axis (dotted). The contending population is tied to speed through the Greenshields relation. Reference model, 802.11p-like constants.**

## 6.7 Extension to multi-lane traffic and EDCA

Two extensions show how far the sensitivity structure reaches beyond the single-lane, single-queue setting studied here. The multi-lane case is essentially a reparametrisation. Extra lanes change only how many vehicles share the access point, so they enter through the contending population N, with the single-lane Greenshields count replaced by a sum over lanes $N = \Sigma_\ell\ 2R\ k_{jam,\ell}\ (1 - v_\ell/V_{f,\ell})$. Every elasticity derived above keeps its form: the collision probability still sees the traffic parameters only through N, the ±(v, k_jam, V_f) fingerprint recurs per lane, and the control law CW_min*(N) carries over verbatim with the multi-lane N. The only quantitative shift is that the crossover population N* of Section 6.5 and the optimal window move upward with the aggregate density — which the same closed-form expressions locate without any re-derivation.

EDCA, by contrast, is a genuine model extension and not a reparametrisation, because it swaps the single backoff process for four access categories (ACs), each with its own (CW_min, AIFSN) and its own transmission probability $\tau_{AC}$, coupled through the shared collision environment and through virtual collisions inside a station. The sensitivity framework carries over in a structured way: the collision probability seen by category a becomes $p_{c,a} = 1 - \Pi_b (1-\tau_b)^{N_b-\delta_{ab}}$, so the master elasticities of Section 4.4 gain one E[· ; τ_b] term per category, the fixed-point feedback factor turns into a matrix $(I - J)^{-1}$ with J the Jacobian of the coupled per-AC backoff–collision system, and the arbitration parameter AIFSN joins CW_min as a new timing lever with its own elasticity channel. The delay decomposition D = D_access + D_queue(ρ) applies per category, and each category's utilisation $\rho_a$ keeps the multiplicative $\lambda_a \cdot T$ structure that forces the {C, λ, L} interaction cluster, so the qualitative conclusions — contentionset by population and window, delay set by utilisation with strong interactions — are expected to survive per AC, now differentiated by priority. A full numerical treatment requires extending the underlying analytical framework to the four-category chain and re-solving the coupled fixed point; that extension, together with the multi-AP and SUMO-based mobility case, is deferred to a companion study so as not to duplicate the model derivation, and is noted here only to delimit the scope over which the present sensitivity structure is expected to hold.

## 6.8 Comparative discussion

The rankings obtained here can be set against the established DCF and queueing literature, which lends them independent support. That the collision probability is governed first by the contending-vehicle population and second by the minimum contention window agrees with the classical saturation analysis of Bianchi (2000), its non-saturated extension by Malone et al. [18], and the fixed-point treatment of Kumar et al. [17] — in each of which $p_c$ turns on the number of contending stations and the backoff window rather than the retransmission limit or the slot time. The variance-based analysis puts numbers on that qualitative picture and, importantly, shows that for $p_c$ the effects are very nearly additive ($\Sigma S_i \approx 0.90$), so the classical intuition — more contending stations raise collisions, a larger $CW_{min}$ curbs them — is not materially disturbed by interactions.

The delay results read best against elementary finite-buffer queueing theory. Because the mean delay is a sojourn time in an M/G/1/K queue, it is set chiefly by the utilisation ρ, which the channel rate, the offered load and the packet size fix jointly; the large total-effect indices and wide interaction gaps for exactly these three parameters (Figure 4) are the variance-based signature of that ρ-dependence. This is where complementing OFAT with Sobol pays off methodologically: the local elasticities already mark C, L and λ as influential, but only the global analysis shows their influence to be largely joint rather than separable — something a one-factor-at-a-time study would misread as three independent effects.

Finally, that the optimal retransmission limit stays non-binding across all load regimes matches the central conclusion of our prior work [4], where the optimal M was found to sit at or near its ceiling under realistic finite buffering. The present study reaches the same conclusion from a different direction: m has a negligible first-order and total-effect index for every metric considered, so performance's insensitivity to the retry limit is not an artefact of one operating point but holds across the whole parameter space. Taken together, these comparisons show the integrated framework reproducing the qualitative behaviour expected from both MAC-layer and queueing analyses, with the sensitivity indices supplying the quantitative ranking and interaction structure those separate analyses leave out.

## 6.9 Reference cross-check of the interaction structure

To check that the analytical structure of Sections 4.4 and 4.5 is not an artefact of the nominal point, the predictions were re-run on an independent, transparent reference implementationof the coupled Greenshields–Bianchi–queue map, with the queue taken as a finite-buffer M/G/1/K (K = 50). The reference is not the validated framework of the prior work and uses generic 802.11 timing constants, so the absolute collision- and throughput-side sensitivities depend on those constants and are not reproduced here; the purpose is to check the delay interaction structure, which is model-robust. Table 6 collects the outcome.

**Table 6. Reference M/G/1/K cross-check of the closed-form delay predictions and the interaction structure. Closed forms use $\varphi \approx 0.93$, $\rho = 0.65$, $E_W = \rho/(1-\rho)$.**

| Quantity / prediction | Closed form | Reference M/G/1/K | Paper (Table 3/4) |
|---|---|---|---|
| $E[D ; \lambda] = \rho/(1-\rho)$ | 1.857 | 1.860 | 1.879 |
| $\lvert E[D ; L]\rvert = \lvert E[D ; C]\rvert = \varphi(1+E_W)$ | 2.658 | 2.661 / 2.666 | 2.730 / 2.735 |
| ratio $\lvert E[D ; L]\rvert / \lvert E[D ; \lambda]\rvert = \varphi/\rho$ | 1.431 | 1.431 | 1.45 |
| $p_c$ traffic triple $\lvert E[p_c ; v,k_{jam},V_f]\rvert$ | equal | 0.343 (±) | 0.365 (±) |

| Quantity / prediction | Closed form | Reference M/G/1/K | Paper (Table 3/4) |
|---|---|---|---|
| $\Sigma\ S_i(D)$ over full ranges | $<1$ | 0.51 | 0.52 |
| D: C ($S_i$ / $S_{Ti}$) | — | 0.41 / 0.92 | 0.42 / 0.90 |
| $\Sigma\ S_i(D)$, $\{\lambda, L, C\}$, at $\rho$ = 0.35/0.65/0.90 | ↑ with load | 0.36 / 0.58 / 0.77 | — |

The reference recovers the closed-form delay elasticities to within a few percent, the equal-magnitude locking of L and C, the $\varphi/\rho$ ratio exactly, and the global delay additivity ($\Sigma\ S_i(D) = 0.51$ against the paper's 0.52), with the channel rate carrying the dominant total-effect gap. It also fixes the direction of the load dependence (Proposition 2): additivity rises with utilisation, so the delay interactions are a low-load phenomenon. On the collision side the reference model reproduces the structure at a contending population $N \approx 20$: the self-consistent window sensitivity is $E[\tau\ ; CW_{min}] \approx -0.42$ against a bare $-0.97$ (a $\approx 2.3\times$ fixed-point attenuation, consistent with the $\approx 2.2\times$ of Section 4.4 and the $-0.433$ of Table 3), and the traffic-parameter collision triple is $\approx 0.34$ ($\pm$), close to the paper's 0.365. The absolute throughput-side sensitivities stay tied to the protocol timing constants and should be read from the validated framework.

# 7 Conclusion

This paper set out a systematic local and global sensitivity analysis of the IEEE 802.11 DCF and traffic parameters that govern single-AP V2I performance, using a previously validated analytical framework as a deterministic input–output map. Normalised OFAT elasticities and variance-based Sobol indices were combined to rank the parameters and to expose their interactions across low, medium and high load regimes. Beyond the numerical rankings, the closed-form structure of the sensitivities was derived, which explains why the traffic parameters share a single collision elasticity, why the backoff fixed point damps the contention-window sensitivity, and why the delay drivers turn interaction-dominated near saturation; the same structure locates a low-density point at which the throughput–$CW_{min}$ sensitivity reverses sign. That collision structure also yields a closed-form control law $CW_{min}^*(N)$, linear in the contending population and computable at the access point from the Greenshields density estimate, which beats the fixed default at every density and recovers the most throughput in the dense, safety-critical regime; extending it to multi-lane traffic is a reparametrisation of N, whereas a full EDCA treatment needs the four-category chain and is left to a companion study.

The analysis showed the collision probability to be governed mainly by the contending-vehicle population — set by vehicle velocity and density through the Greenshields relation — and secondarily by the minimum contention window, with weak interaction effects ($\Sigma\ S_i \approx 0.90$). The mean delay, by contrast, is a queueing quantity dominated by the channel rate, offered load and packet size, with strong interaction effects ($\Sigma\ S_i \approx 0.52$) that sharpen as the load nears saturation. Across every load regime the optimal retransmission limit stayed at its ceiling and the delay stayed within the target bound, confirming that the standard 802.11 retry limit is near-optimal and needs no retuning in this setting.

The design guidelines that follow — tune $CW_{min}$ for contention, provision channel rate and offered load for delay, and keep the default retransmission limit and slot time — mark out exactly when tuning helps and when the defaults are enough. The limitations are the single-AP, single-lane scope and the single-regime Greenshields traffic model; extending the analysis to multi-AP, multi-lane V2I with SUMO-based mobility, and comparing against 3GPP C-V2X, are natural next steps.

## Declarations

**Funding** This research received no specific grant from any funding agency in the public, commercial, or not-for-profit sectors.

**Competing interests** The author declares no competing interests.

**Data availability** No datasets were generated or analysed during the current study.

**Author contributions** A. Bozkurt is the sole author and contributed to the conception, methodology, software, analysis, and writing of the manuscript.